\documentclass{aa}
\usepackage{graphicx}
\usepackage{txfonts} 

\newcommand{\acap}{\noindent}
\newcommand{\msun}{\ensuremath{M_\odot}}
\newcommand{\rsun}{\ensuremath{R_\odot}}

\newcommand{\dif}{\ensuremath{\mathrm d}}
\newcommand{\azero}{\ensuremath{A_0}}

\newcommand{\ai}{\ensuremath{A_{\rm i}}}
\newcommand{\af}{\ensuremath{A_{\rm f}}}
\newcommand{\aff}{\ensuremath{A}}
\newcommand{\affn}{\ensuremath{A_{\rm n}}}
\newcommand{\affx}{\ensuremath{A_{\rm x}}}

\newcommand{\afai}{\ensuremath{A_{\rm f}/A_{\rm i}}}
\newcommand{\affai}{\ensuremath{A/A_0}}

\newcommand{\ace}{\ensuremath{\alpha_{\rm ce}}}
\newcommand{\betaa}{\ensuremath{\beta_{\rm a}}}
\newcommand{\betaat}{\ensuremath{\tilde{\beta}_{\rm a}}}
\newcommand{\betag}{\ensuremath{\beta_{\rm g}}}
\newcommand{\gammace}{\ensuremath{\gamma_{\rm CE}}}
\newcommand{\fen}{\ensuremath{f_{\rm en}}}

\newcommand{\msec}{\ensuremath{m_{2}}}
\newcommand{\msece}{\ensuremath{m_{\rm 2,e}}}
\newcommand{\msecc}{\ensuremath{m_{\rm 2,c}}}
\newcommand{\mpri}{\ensuremath{m_{1}}}
\newcommand{\mprii}{\ensuremath{m_{\rm 1,i}}}

\newcommand{\mprin}{\ensuremath{m_{\rm 1,n}}}

\newcommand{\mpwd}{\ensuremath{m_{\rm 1R}}}
\newcommand{\mswd}{\ensuremath{m_{\rm 2R}}}
\newcommand{\mdd}{\ensuremath{M_{\rm DD}}}
\newcommand{\mddt}{\ensuremath{\tilde{M}_{\rm DD}}}

\newcommand{\mddn}{\ensuremath{M_{\rm DD,n}}}
\newcommand{\mddx}{\ensuremath{M_{\rm DD,x}}}

\newcommand{\md}{\ensuremath{m^{\rm d}}}
\newcommand{\mdi}{\ensuremath{m^{\rm d}_{\rm i}}}
\newcommand{\mdf}{\ensuremath{m^{\rm d}_{\rm f}}}

\newcommand{\mb}{\ensuremath{M_{\rm b}}}
\newcommand{\mbi}{\ensuremath{M_{\rm b,i}}}
\newcommand{\mbs}{\ensuremath{M_{\rm b,s}}}

\newcommand{\mwdn}{\ensuremath{m_{\rm WD,n}}}

\newcommand{\Mstar}{\ensuremath{\mathcal{M}}}
\newcommand{\mol}{\ensuremath{\mathcal{M}/L_{\rm B}}}

\newcommand{\fia}{\ensuremath{f_{\rm Ia}}}
\newcommand{\fiasd}{\ensuremath{f_{\rm Ia}^{\rm SD}}}
\newcommand{\fiadd}{\ensuremath{f_{\rm Ia}^{\rm DD}}}
\newcommand{\fiaddw}{\ensuremath{f_{\rm Ia}^{\rm DDW}}}

\newcommand{\kalpha}{\ensuremath{k_{\alpha}}}
\newcommand{\tburst}{\ensuremath{\Delta t_{\rm B}}}
\newcommand{\Mburst}{\ensuremath{\mathcal{M}_{\rm B}}}

\newcommand{\psizero}{\ensuremath{\psi_{0}}}
\newcommand{\deltat} {\ensuremath{\Delta t}}

\newcommand{\tage}{\ensuremath{\tau}}
\newcommand{\taui}{\ensuremath{\tau_{\rm i}}}
\newcommand{\taun}{\ensuremath{\tau_{\rm n}}}
\newcommand{\taums}{\ensuremath{\tau_{\rm MS}}}
\newcommand{\taux}{\ensuremath{\tau_{\rm x}}}

\newcommand{\taugr}{\ensuremath{\tau_{\rm gw}}}
\newcommand{\tauni}{\ensuremath{\tau_{\rm n,i}}}
\newcommand{\tinf}{\ensuremath{\tau_{\rm n,inf}}}
\newcommand{\taunx}{\ensuremath{\tau_{\rm n,x}}}
\newcommand{\taugi}{\ensuremath{\tau_{\rm gw,i}}}
\newcommand{\taugx}{\ensuremath{\tau_{\rm gw,x}}}
\newcommand{\taust}{\ensuremath{\tau_{\rm n}^{\star}}}

\newcommand{\fmr}{\ensuremath{f}}
\newcommand{\fmrx}{\ensuremath{f_{\rm max}}}
\newcommand{\fni}{\ensuremath{f^{\rm W}_{1}}}
\newcommand{\fnii}{\ensuremath{f^{\rm W}_{2}}}
\newcommand{\fniii}{\ensuremath{f^{\rm W}_{1,2}}}
\newcommand{\fddc}{\ensuremath{f^{\rm C}}}

\newcommand{\stepc}{\ensuremath{S^{\rm C}}}
\newcommand{\stepw}{\ensuremath{S^{\rm W}}}

\newcommand{\aia}{\ensuremath{A_{\rm Ia}}}

\newcommand{\avefiab}{\ensuremath{\langle f_{\rm Ia}
\rangle_{t-\Delta t,t}}}
\newcommand{\avefiat}{\ensuremath{\langle f_{\rm Ia}\rangle_{\taui,t}}}
\newcommand{\avefiapsi}{\ensuremath{\langle f_{\rm Ia}\rangle_{\psi(t)}}}
\newcommand{\rsnu}{\ensuremath{\mathcal{R}_{\rm SNu}}}

\begin{document}

\title{The rates of Type Ia Supernovae. I. Analytical Formulations.}

\author{Laura Greggio\inst{1}}

\offprints{L. Greggio}

\institute{INAF, Osservatorio Astronomico di Padova, vicolo 
dell'Osservatorio 5, 35122 Padova, Italy\\
\email{greggio@pd.astro.it}
   }
 
\date{Received ; accepted }

\abstract{ 
The aim of this paper is to provide a handy tool to compute the impact
of Type Ia SN (SNIa) events on the evolution of stellar systems. 
An effective formalism to couple the rate of SNIa explosions from a single
burst of star formation and the star formation history is presented, which
rests upon the definition of the realization probability of the SNIa event
(\aia) and the distribution function of the delay times (\fia(\tage)).
It is shown that the current SNIa rate in late type galaxies constrains 
\aia\ to be on the order of
$10^{-3}$ (i.e. 1 SNIa every 1000 \msun\ of gas turned into stars), while
the comparison of the current rates in early and late type galaxies 
implies that \fia\ ought to be more populated at short delays.
The paper presents analytical formulations for the description of the
\fia\ function for the most 
popular models of SNIa progenitors, namely Single Degenerates 
(Chandrasekhar and Sub--Chandrasekhar exploders), and Double Degenerates.  
These formulations follow entirely from 
general considerations on the evolutionary behavior of stars in
binary systems, modulo a schematization of the outcome of the phases of
mass exchange,
and compare well with the results of population synthesis codes, for the
same choice of parameters.
The derivation presented here offers an immediate astrophysical
interpretation of the shape of the \fia\ functions,
and have a built in parametrization of the key properties of the alternative
candidates. The important parameters appear to be
the minimum and maximum masses of the components of the binary systems
giving rise to a SNIa explosions, the distribution of the primary mass 
and of the mass ratios in these systems, the distribution of the separations
of the DD systems at their birth.
The various models for the progenitors correspond to markedly different 
impact on the large scales;
correspondingly, the model for the progenitor can be 
constrained by examining the relevant observations.  
Among these, the paper concentrates on the trend of the current 
SNIa rate with parent galaxy type. The recent data by Mannucci et al.
(\cite{Mannu}) 
favor the DD channel over the SD one, which tends to predict a too steep
distribution function of the delay times. The SD scenario can be 
reconciled with the observations only if the distribution of the mass 
ratios in the primordial binaries is flat and the accretion efficiency
onto the WD is close to 100\%. 
The various models are characterized by different timescales for the
Fe release from a single burst stellar population. 
In particular the delay time within which half of the SNIa
events from such a population have occurred, ranges between 0.3 and 3
Gyr, for a wide variety of hypothesis on the progenitors.

         }

\maketitle

\section{Introduction}



The evolution of the rate of Type Ia Supernovae (SNIa) with time 
is a fundamental ingredient for the study of a variety of astrophysical
issues, ranging from the chemical evolution of stellar systems, to the
interpretation of the SNIa rates as a function of redshift.


 
Indeed, while type II Supernovae (SNII) precursors are short lived, massive 
stars, so that their rate evolves (almost) in pace with the rate of star 
formation,
SNIa come from binary systems with a wide range of lifetimes, as indicated by 
their occurrence in both late and early type galaxies. 
Thus, the SNIa products are released to the 
interstellar medium over longer timescales, compared to SNII products.
Since the $\alpha$ elements are mostly produced by SNII, while SNIa are
important contributors of iron, the shorter the formation timescale of a 
stellar
system is, the higher the $\alpha$ to Fe abundance ratio recorded in its stars.
This argument is at the basis of the evaluation of the formation timescales of 
stellar systems from the $\alpha$ to Fe abundance ratios, and has 
been used to estimate the formation timescale of the halo of our galaxy from 
individual stellar abundances (e.g. Matteucci \& Greggio \cite{MG86}), 
as well as to infer short formation timescales for Es from the Magnesium
and Iron indices in their spectra (e.g. 
Matteucci \cite{Chicchi94}; Greggio \cite{IO97}; Thomas et al.
\cite{Daniel05}).
SNIas are also thought to produce a major fraction of the iron in the
intracluster medium in galaxy clusters 
(e.g. Matteucci \& Vettolani \cite{MV88}; Renzini \cite{Alvio97}); 
the temporal behavior of the SNIa rate will
then impact on the iron abundance of the intracluster gas at high redshift.
 
Type Ia SNe provide the heating mechanism of the mass lost by stars in 
Ellipticals, an therefore determine the dynamical evolution of the gas 
in these galaxies. Following Renzini (\cite{Alvio96}), this evolution 
depends on the 
balance between the rate of mass return, and the secular evolution of the 
SNIa rate: if the latter decreases faster than the former, 
the early stages are characterized by 
supersonic winds, which then turn to subsonic outflows, and eventually 
to inflows (Ciotti et al. \cite{Ciotti}). The reverse sequence
applies instead in the case of a mild secular evolution of the SNIa rate,
with early inflows eventually turning into winds, such as in the models of 
Loewenstein \& Mathews (\cite{Mathews87}).


Finally, cosmological applications of the SNIa rate include the possibility
of deriving clues on the SNIa progenitors, 
and/or constraining the cosmic star formation history 
from the evolution of the SNIa rate with redshift
(Madau et al. \cite{Piero98}; Strolger et al.\cite{Strol04}).
Understanding the evolutionary path which leads to
a Type Ia explosion is of great importance to assess the use of SNIa
as distance indicators, and to derive the cosmological parameters (Riess
et al. \cite{Riess}; Perlmutter et al. \cite{Perl99}).


To address these issues and quantitatively interpret the related observations
we need a suitable description of the evolution of the SNIa rate
from a single burst stellar population. From a theoretical point of view,
this rate is difficult to derive, first because the nature of the 
progenitors of SNIa events is still an open question, and, second, because
any theoretical rendition is highly model dependent.
While in the literature there is a general consensus that SNIa originate form 
the thermonuclear explosion of Carbon and Oxygen (CO) White Dwarfs (WD), 
various evolutionary paths may lead to such event. 
Common to all the models is the first part of the evolution, dealing with
a close binary system with a primary component less massive than $\sim$ 
8 \msun, so that it evolves into a CO WD. When the secondary component expands
and fills its Roche Lobe, the primary may or may not accrete the matter shed
by its companion. If the accretion rate is approximately $\sim 10^{-7}$\msun/yr
(Nomoto \cite{Ken82}), the accreted matter burns on top of 
the WD, the object remains confined within its Roche Lobe and grows in mass 
(Whelan \& Iben \cite{WI73}).
However, Iben \& Tutukov (\cite{IT84}) pointed out that in most cases the 
secondary expands at such a high rate that the accretion rate 
exceeds the mentioned limit, implying that a common 
envelope (CE) forms around the two stars. Orbiting inside the CE the two cores 
spiral in, and orbital energy is transferred to the envelope 
which is eventually lost. The system emerges form the CE phase as a 
close double WD, which will  
merge due to the emission of gravitational wave radiation. 
Another interesting possibility is that, when subject to a large accretion 
rate, the WD develops a strong radiative 
wind, to the effect of stabilizing the mass transfer, thus allowing 
the WD to grow in mass, and eventually explode 
(Hachisu, Kato \& Nomoto \cite{HKN96}).
Within all scenarios explosion occurs either when the CO WD reaches the 
Chandrasekhar mass and Carbon deflagrates at the center (Chandra exploders), 
or when a massive enough helium layer is accumulated on top of the CO WD, so 
that Helium detonates, inducing off center Carbon detonation (e.g. Woosley 
\& Weaver \cite{Stan94}) before the Chandrasekhar mass is 
reached (Sub--Chandra exploders).

Different arguments can be found in favor or against both scenarios
(e.g. Livio \cite{Livio01}), 
generally referred to as Single Degenerate (SD) and Double Degenerate (DD),
depending on whether the SNIa precursor is a system with one or two WDs.
Briefly, the SD model is supported by the observational detection of several 
classes of objects that can be considered as potential SNIa precursors
of the SD variety, i.e. Cataclysmic Variables, Symbiotic Stars, and 
Supersoft X-Ray Sources (Munari \& Renzini \cite{Ulisse}; Kenyon et al. 
\cite{Kenyon}; Van den Heuvel et al. \cite{VdH}; Rappaport, Di Stefano \&
Smith \cite{Rappaport}). Additional support to the Cataclysmic
Binaries channel came recently from the detection of a candidate 
companion to Tycho's supernova (Ruiz-Lapuente et al. \cite{Pilar04}).
On the other hand,
the fine tuning of the mass accretion rate limits considerably the volume
in the parameter space for a successful SNIa explosion in the SD model.
As a consequence, it seems likely that only a small
fraction of events can be realized through this channels in our galaxy 
(Fedorova, Tutukov \& Yungelson \cite{Fedorova}; 
Han \& Podsiadlowski \cite{Han04}, but see 
Hachisu, Kato \& Nomoto \cite{HKN99} for a different point of view).

Several attempts have been made to establish the binary frequency among
White Dwarfs, and to determine the distribution of total masses and 
periods of the binary systems, in order to assess the likelihood of
the DD channel as SNIa progenitor (Robinson \& Shafter \cite{Robin};
Bragaglia et al. \cite{Angie}; Foss, Wade \& Green \cite{Foss}; Saffer,
Livio \& Yungelson \cite{Saffer}; Maxted \& Marsh \cite{Marsh}).
To date, the most comprehensive effort to find SNIa precursors among 
DD systems is the SPY project (Napiwotzki et al. \cite{SPY}), whose
results have been recently summarized in Napiwotzki et al. (\cite{SPYresults}):
many close DD systems have been found, with one very good candidate SNIa
precursor, with a total mass exceeding the Chandrasekhar limit and expected 
to merge within a Hubble time. In addition, a few other systems come close to 
qualify as SNIa precursors. In general, it seems that the masses and
period distributions of the binary WDs confirm the prediction of the
population synthesis models; according to these models the DD evolutionary 
channel 
{\em is} able to provide enough merging events to match the 
current SNIa rate measured in our galaxy, which is similar to the
typical SNIa rate in Spirals. On the other hand, 
theoretical calculations show 
that the merging of two massive WDs may lead to accretion induced collapse, 
rather than to SNIa explosion (Saio \& Nomoto \cite{Sano}), so that the
ultimate fate of these candidates may be a neutron star. 

The various models for the progenitors, SD or DD, undergoing
Chandra or Sub--Chandra explosions, correspond to rather different 
temporal behavior of the SNIa rate (see e.g. Fig. 2 in Yungelson 
\& Livio \cite{YL}, hereafter YL). 
In the current literature there are several examples of theoretical 
computations of the SNIa rate performed with
population synthesis codes: starting from a 
primordial distributions of binary masses, mass ratios and separations,
the computations 
follow the evolution of the stellar systems under some prescriptions for the 
mass exchange between the binary components, to determine the final outcome
(Tutukov \& Yungelson \cite{Tuyu}; Yungelson et al. \cite{Yu94}; 
Ruiz-Lapuente, Burkert \& Canal \cite{Pilar95}; Han, Podsiadlowski \& 
Eggleton \cite{Han95}; Nelemans et al. \cite{Nele01}; 
De Donder \& Vanbeveren \cite{Deny}).
The results of these simulations depend on a variety of input 
parameters and assumptions, whose
role is difficult to gauge, so that they are not suited to easily explore the 
parameter space for the SNIa progenitors' models.
In addition, the population synthesis codes yield  
numerical outputs, which are difficult to incorporate in codes which
follow the evolution of galaxies or of galaxy clusters. Indeed, 
most of the astrophysical applications of the SNIa rates in the literature 
are based either on the analytical formulation by Greggio \& Renzini
(\cite{IO83}), or on the parametrization proposed by Madau et al. 
(\cite{Piero98}).
However, the former is derived only in the framework of the SD model; the 
latter is a convenient mathematical expression, but it is only marginally 
related to the physics of stellar evolution. 

In this paper I provide relatively simple analytical formulation 
for the SNIa rate, which allows us to identify the most critical parameters 
and should help restrict the choice among the candidate precursors.
Both the SD (Chandra and Sub--Chandra) and the DD (only Chandra) models
are considered, so as to provide a handy way for investigating on the 
impact of the different SNIa models on astrophysical issues for which the
SNIa rate is important. 
Section 2 presents a coherent formalism to couple a particular 
SNIa model with the star formation history of a system. The analytic
expressions for the SNIa rate for the SD and DD models are derived
in Sections 3 and 4 respectively.
Readers mostly interested in the main results may skip these sections, since
a general description of the analytic \fia\ function appears at the beginning
of Section 5, where they are compared to  
the predictions of population synthesis codes. In addition, Section 5 presents
an attempt to constrain the SNIa progenitors from the systematic
trend of the SNIa rates with galaxy type.
Finally, some concluding remarks appear in Section 6.
The mathematics used to derive of the analytic relation for the DD model 
is (mostly) described in the Appendix, for an easier readability of the text.

\section{Formalism}

A convenient formulation of the SNIa rate to follow the evolution of
stellar systems rests upon the definition of the distribution
function of the delay times, i.e. the time elapsed between the birth of 
a SNIa progenitor and its explosion. I indicate this function with
\fia(\tage), defined in the range (\taui , \taux) , respectively
the minimum and maximum possible delay times, and consider \fia(\tage)
normalized to 1: $\int_{\taui}^{\taux}\fia(\tage) \cdot \dif \tage = 1 $.
The minimum delay time \taui\ is the minimum evolutionary lifetime of the 
SNIa precursors: for the SD model this is the nuclear lifetime of the most
massive stars which produce a WD, that is an $\sim$ 8 \msun\ star, which
evolves off the MS in $\sim 0.04$ Gyr. For the DD model, \taui\ could be
appreciably larger than this because of the additional gravitational delay, 
i.e. the time taken by the DD system to merge due to the gravitational 
wave radiation. 
The maximum delay time \taux\ is quite sensitive to the model for the SNIa
precursor, as will be seen later. At this point I just notice that, if
elliptical galaxies formed the bulk of their stars in a short initial burst,
the maximum delay time of their inhabiting SNIa precursors must be 
on the order of a Hubble time, or more.

At a given epoch $t$, the contribution to the SNIa rate 
from progenitors with delay times in the range (\tage , \tage+d\tage) is

\begin{equation}
 \dif \dot{n}_{Ia} = \dot{n}_\star (t-\tage) \times \aia(t-\tage) \times 
 \fia (\tage)\,\dif \tage
\label{eq_rate0}
\end{equation}

\acap
where $\dot{n}_\star (t-\tau)$ is the birth rate at 
epoch $(t-\tage)$, and \aia\ is the realization probability 
of the SNIa scenario from the stellar generation born at $(t-\tage)$ 
\footnote {If $N_{\star}$ is the number of stars born 
at epoch (t-\tage), \aia $\times$ $N_{\star}$ is the number of SNIa 
events produced by this stellar generation ever.}.
In Eq.~(\ref{eq_rate0}) I have considered the possibility of variations of 
\aia\ during the galaxy evolution. For example, one could expect
larger realization probabilities of the SNIa channel at higher metallicities
because, during their evolution, stars expand to larger 
radii (Greggio \& Renzini \cite{IO90}), hence have a better chance to 
fill their 
Roche Lobe. In addition, at high metallicity the accretion onto the 
pre-supernova WD could be more efficient 
(Hachisu et al. \cite{HKN96}), also implying larger values for \aia. 

Following Tinsley (\cite{Tinsley}) notation:

\begin{equation}
\dot{n}_\star (t-\tau)=\psi(t-\tage) \times \int_{m_i}^{m_s} {\phi(m)\,\dif m}
\end{equation}

\acap
where $\phi(m)$ is the initial mass function (IMF) by number,
and $m_i$ and $m_s$ are the lower and upper mass limits. As usual, I adopt 
a power law IMF, with total mass normalized to 1, so that the star formation
rate (SFR) $\psi$ 
is the mass that goes into stars per unit time. It follows:

\begin{equation}
\dot{n_\star}(t-\tau)=\psi(t-\tage) \times \kalpha
\label{eq_tinsley}
\end{equation}

\acap
where $\kalpha$, which is the number of stars per unit mass in one stellar 
generation, depends on the IMF. For example, for
a mass distribution $\phi(m) \propto m^{-\alpha}$ ranging from 0.1 to 120 
\msun, $\kalpha$ is equal to 2.83 and 1.55, respectively for Salpeter 
($\alpha=2.35$) and Kroupa (\cite{Pavel01}) 
($\alpha=2.3$ in $m\geq 0.5$\msun\ and $\alpha=1.3$ in $m\leq 0.5$\msun) IMFs.

By substituting Eq. (\ref{eq_tinsley}) into Eq. (\ref{eq_rate0}), and summing
over all the contributions from the past stellar generations,
the SNIa rate at epoch $t$ is:

\begin{equation}
\dot{n}_{Ia}(t) = \kalpha \times \int_{\taui}^{\min(t,\taux)}
{\psi(t-\tage) \,\, \aia(t-\tage) \,\, \fia(\tage)\,\dif \tage}.
\label {eq_rate}
\end{equation}

This equation allows one to insert consistently the SNIa events in codes
which describe the evolution of galaxies, once the distribution of the delay
times \fia\ and the fraction \aia\ are specified.
Notice that these are the only results of the modeling of binary populations
of SNIa precursors 
which impact on the evolution of stellar systems. 
In other words, for the astrophysical applications,
the particular prescriptions used in the population synthesis codes are of 
limited interest, while most important is the distribution function of the 
delay times and the total realization probability of the SNIa scenario out of
one stellar generation. 

Eq. (\ref{eq_rate}) can be easily specified for a single generation of
stars by considering a star formation episode started at $t=0$ and proceeded
at a constant rate $\psi_{\rm B}$ for a very short time (\tburst).
In this case, the integrand function is non zero only in a narrow age range 
around $\tage = t$ so that:

\begin{eqnarray}
\dot{n}_{Ia}(t) & = & \kalpha \cdot \psi_{\rm B} \cdot \tburst \cdot
\aia(t=0) \cdot \fia(\tage=t) \nonumber \\
 & = & \kalpha \cdot \Mburst \cdot \aia \cdot \fia(t)
\label {eq_rateb}  
\end{eqnarray}

\acap
where \Mburst\ is the mass that went into stars in the burst.
Thus {\em  the SNIa rate following an instantaneous burst of star formation is
proportional to the distribution function of the delay times} through a factor
which is the product of the total stellar mass formed
in the burst, times the realization probability of the SNIa scenario, times
the number of stars per unit mass, characteristic of the IMF.

\subsection {Observational constraints}

Eq.(\ref{eq_rate}) shows that the SNIa rate results from the convolution
of the distribution function of the delay times and the SF history: in order
to derive clues on the former from the observed rates
I consider a family of models, characterized by a constant SFR
(\psizero) starting at $t=0$, and with variable durations (\deltat).
I also assume that early (late) type galaxies are represented by models 
with short (long) \deltat.
Although the SF history in real galaxies is much more complex than that,
this schematic description leads to interesting relations between the
observed SNIa rates and the key quantities \aia\ and \fia, which will not be
far from those that one can obtain with a thorough modeling of the 
galaxy evolution.
 
For this family of models, the minimum delay time which contributes to the
rate at epoch $t$ is either \taui\ (i.e. the absolute minimum delay of the
SNIa progenitors), or the time elapsed from the end of
the burst: ~$t-\deltat$. Taking this into account, and neglecting the 
temporal dependence of the factor \aia, Eq. (\ref{eq_rate}) becomes: 

\begin{equation}
\dot{n}_{Ia}(t) = {\kalpha} \cdot \aia \cdot \psizero \cdot 
\int_{\max(\taui,t-\deltat)}^{\min(t,\taux)} {\fia(\tage)\,\dif \tage}.
\label {eq_ratem}
\end{equation}

%

The two options for the lower integration limit correspond to
two different regimes:

\begin{enumerate}
\item when $\taui > t-\deltat$ , i.e. $t < \taui+\deltat$, we basically map 
the SNIa rate while the SF is still ongoing;

\item when  $\taui < t-\deltat$, i.e. $t > \taui+\deltat$, we get the 
SNIa rate after the burst is completed. 
\end{enumerate}
The transition between regime 1) and 2) occurs at older ages when 
the SF episode lasts longer (i.e. when \deltat\ is larger).
For a currently ongoing SF the transition has not yet occurred. 
 
In regime 1), relevant for late type galaxies, the SNIa rate is given by: 

\begin{equation}
\dot{n}_{Ia}^{L}(t) = \kalpha \cdot \aia \cdot \psizero \cdot 
\int_{\taui}^{\min(t,\taux)} {\fia(\tage)\,\dif \tage}
\label{eq_ratel0}
\end{equation}

showing that it increases with time, as systems with increasingly longer
delays contribute to the explosions, up to $t=\taux$. From then
on the SNIa rate stays constant, and equal to 

\begin{equation}
\dot{n}_{Ia}^{L}(t) = \kalpha \cdot \aia \cdot \psizero
\label{eq_ratel} 
\end{equation}

\acap
since \fia\ is normalized to 1. Notice that this expression is valid only
when $\deltat \geq \taux$, that is when the SF episode lasts long enough to
include all the possible delay times. In this case the current SNIa rate gives 
indications on the realization probability \aia.
For example, Cappellaro, Evans \& Turatto (\cite{Capp99}) report an 
observed SNIa
rate in late type galaxies of $\simeq$ 0.2 SNu \footnote{1 SNu is one event
per century per $10^{10}L_{\rm B,\odot}$.} 
, i.e. $0.2\cdot (L_{\rm B}/L_{\rm B,\odot})\cdot 10^{-12}$ events per year.
Then, approximating \psizero\ with the ratio between the galaxy's stellar mass 
(\Mstar) and its age ($t$) Eq. (\ref{eq_ratel}) yields:

\begin{equation}
\aia \sim 10^{-3} \times \frac{0.2}{\kalpha} \times 
\left(\frac {\Mstar}{L_{\rm B}}\right)^{-1}
\times t_{\rm Gyr}
\label {eq_aia}
\end{equation} 

\acap 
with $L_{\rm B}$ in solar units. Notice that the mass \Mstar\ in
Eq.~({\ref{eq_aia}}) corresponds to the integrated star formation rate, and  
then it differs from the actual galaxy stellar mass because of the gas 
returned by the stars as they evolve. Using Maraston 
(\cite{Cla98},\cite{Cla05}) simple 
stellar population models with solar metallicity, I find that for a constant
SFR the \mol\ ratio at an age of 12 Gyr is $\sim$ 2 or 1.5, respectively 
for a Salpeter and Kroupa IMF. By inserting these values into
Eq. (\ref{eq_aia}) coupled with the appropriate \kalpha, I get
$\aia \sim 5 \times 10^{-4}$ or $10^{-3}$, respectively for Salpeter and 
Kroupa IMF. Thus, one SNIa event is produced for every 2000 (1000) \msun\
of stars formed with a Salpeter (Kroupa) initial mass distribution.
Although this is only a rough estimate, it appears that 
the current rates measured in late type systems require relatively small 
realization probabilities of the SNIa channel. For comparison, the realization 
probability of SNII, estimated as the fraction of stars more massive than
8 \msun\ in a stellar population, is $\sim$ 0.003, 0.007 
respectively for Salpeter and Kroupa IMF. Therefore, out of a stellar
population of 1000 \msun\ roughly 1 SNIa, and 7 SNII are produced.

Considering now regime 2), relevant for early type galaxies, 
as long as $t < \taux$ Eq. (\ref{eq_ratem}) yields:  

\begin{eqnarray}
\dot{n}_{Ia}^{E}(t) & = & \kalpha \cdot \aia \cdot \psizero \cdot \deltat
\cdot \avefiab \nonumber \\
 & = & \kalpha \cdot \aia \cdot \Mstar \cdot \avefiab 
\label{eq_ratee}
\end{eqnarray}

\acap
where \avefiab\ is the average of the distribution function of the delay
times in the range $(t-\deltat;t)$.

After the burst is completed, the SNIa rate scales according to the
\fia\ function, and goes to zero at $t=\taux+\deltat$.
Early type galaxies are currently in this age regime, so that their 
SNIa
rate yields information on the value of the distribution function of
the delay times at ages in the vicinity of the galactic age.
This constraint is better illustrated by constructing
the ratio of the SNIa rate in early and late
type systems at the current epoch. Dividing  Eq. (\ref{eq_ratee}) by  
Eq. (\ref{eq_ratel0}), and considering the rates in SNu, we get:  

\begin{equation}
\rsnu = \frac{{\dot{n}^E}_{Ia,SNu}}{{\dot{n}^L}_{Ia,SNu}} \simeq 
\frac {(\Mstar/L_{\rm B})^E} {(\Mstar/L_{\rm B})^L} \times \frac {\avefiab} {\avefiat},
\end{equation}

\acap
having assumed that the IMF and the \aia\ factors are the same in early and
late type galaxies. According to Cappellaro et al. (\cite{Capp99}),
the current epoch SNIa rate in early and late type systems is 
the same. This puts a constraint on the ratio between
the average values of the function \fia :

\[\frac{\avefiab}{\avefiat} \simeq \frac{(\Mstar/L_{\rm B})^L}{(\Mstar/L_{\rm B})^E}.\] 

Maraston (\cite{Cla98},\cite{Cla05}) models of simple stellar populations at an 
age of 12 Gyr and solar metallicity have 
\mol $\sim$ 13,10 respectively for Salpeter and Kroupa IMF, again having 
considered the mass initially transformed into stars \footnote {Notice that
at 12 Gyr, $\sim$ 30 \% of the initial mass has been returned to the 
interstellar medium by the evolving stars, for a Salpeter IMF. 
The return fraction is
instead 45 \% if Kroupa IMF applies. Therefore, the ratio between the current
stellar mass and the blue luminosity predicted by the models is 
$\sim$ 9, 5 for the two IMFs.}. Combining this estimate of 
the \mol\ ratio for early type systems with the
values quoted above for the late type systems, it turns out that
the current SNIa rates indicate:

\[\frac{\avefiab}{\avefiat} \sim  0.15 .\]

Thus the value of \fia\ at late delay times is substantially 
smaller than its
average value over the whole range (\taui ; $t$); in other words,
{\em the distribution function of the delay times decreases with time}.
This means that the majority of SNIa precursors are relatively short lived.

It is worth emphasizing that if the distribution of the delay times were
flat (i.e. if young and old systems were equally efficient in producing
SNIa events), the observed ratio \rsnu\ would be of the order of 5--10, given
the ratio of the \mol\ values for early and late type galaxies. Therefore,
the fact that the \fia\ function must be decreasing with increasing
delay time is a very robust conclusion.


\subsection {Stellar evolution predictions}

The distribution function of the delay times and the realization
probability of the SNIa channel from a stellar generation can be derived 
from the theory of the evolution of binary systems.
As mentioned in the Introduction, in the literature there are several examples 
of the SNIa rate in stellar systems computed with population
synthesis techniques (see Yungelson \cite{Lev05}). Typically, these models 
predict a \fia\ function characterized by an early maximum, and
a late epoch decline, while the realization probability is indeed on the 
order of $10^{-3}$.
However, these results depend on the adopted
parameters of the simulations, like the star formation history,
the distribution functions of the 
separations and masses of the primordial binaries, and on the specific
prescriptions for the evolution during the hydrodynamical phases of 
the mass transfer. Therefore, (i) the resulting \fia\ functions are model
dependent; (ii) the role of the various input parameters on the output is far
from straightforward.

On the other hand, on general grounds, Ciotti et al. (\cite{Ciotti}) give a 
motivation for the late epoch decline related to the temporal behavior of the 
clock of the explosions. Indeed, some general considerations can be made 
which strongly characterize the shape of the distribution function of the 
delay times, as I show in the following sections. 
I consider separately  the two main categories of SD and DD progenitors,
and derive analytical formulations for the \fia(\tage) function, in
the attempt of clarifying the role of the important stellar evolution 
parameters.

\section {Single Degenerates}

In the SD model, the clock of the supernova event is set by the 
evolutionary lifetime of the secondary. A fit to Girardi et al. 
(\cite{Girardi00}) solar metallicity tracks yields the following
relation between stellar mass (in \msun) and Main Sequence (MS) lifetime 
(\taums, in years ), 
valid in the range ($0.8 \lesssim \msec \lesssim 8)\msun$,
which corresponds to $(0.04 \lesssim \taums \lesssim 25)$ Gyr :

\begin{equation}
\log\,\msec = 0.0471 \times (\log\,\taums)^2 - 1.2 \times \log\,\taums +7.3 .
\label{eq_mto}
\end{equation}

\acap
The total delay time is  basically equal to the MS lifetime of
the secondary component of the binary system, the Post--MS 
phase being in any case much 
shorter than the hydrogen burning lifetime. Therefore,
for one stellar generation, the number
of explosions within (\tage , \tage +d\tage) is proportional to the number
of binaries with secondary mass between \msec\ and \msec+d\msec\ such that the
evolutionary lifetime of \msec\ is \tage:

\begin{equation}
\fiasd(\tage)\,\,|\dif \tage| \propto n(\msec)\,\,|\dif \msec|,
\end{equation}

\acap
which implies:

\begin{equation}
\fiasd(\tage) \propto n(\msec) \cdot |\dot{\msec}|.
\label{eq_fiasd}
\end{equation}


\begin{figure}
\resizebox{\hsize}{!}{\includegraphics{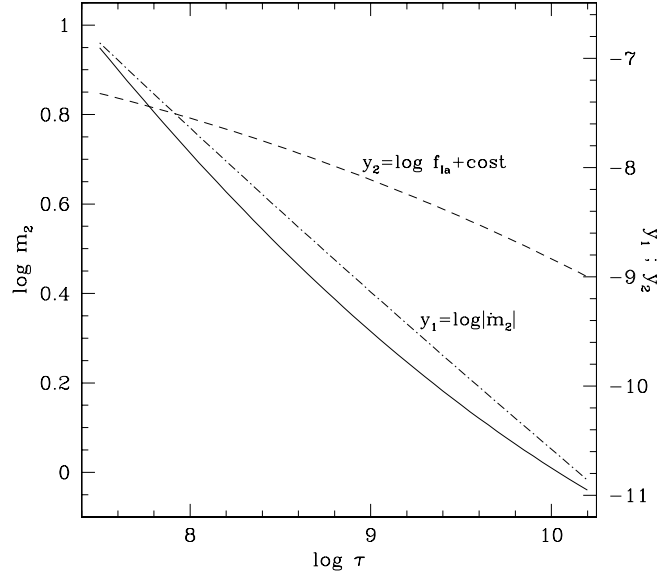}}
\caption{Evolutionary mass (solid line) and its derivative
(dot-dashed line) as a function of the MS lifetime (in years), 
for solar metallicity tracks. The derivative (to be read on the right axis) 
drops by $\sim$ 4 orders of magnitude as the MS lifetime 
increases from $\sim$ 40 Myr to $\sim$ 20 Gyr. The dashed line represents the
distribution function of the delay times if the secondaries follow 
a Sapeter IMF (see text).}
\label{fig_mto}
\end{figure}

Figure ~\ref{fig_mto} shows the evolutionary mass from Eq.~(\ref{eq_mto}) and 
its derivative as a function of the delay time \tage = \taums. 
The derivative is very well approximated by the power law 
$|\dot{\msec}| \propto \tage^{-1.44}$ over the whole range 
from 40 Myr to $\gtrsim$ 15 Gyr. 
In Eq. (\ref{eq_fiasd}), as \tage\ increases, the factor $|\dot{\msec}|$ 
strongly decreases. However, longer delay times
correspond to smaller evolutionary masses \msec,
and the final shape of the distribution function 
depends on the distribution of secondary masses as well. 
The dashed
line in Fig. \ref{fig_mto} shows the result obtained with a 
Salpeter distribution $n(\msec) \propto \msec^{-2.35}$, clearly not
steep enough to imply a \fiasd(\tage) increasing at 
late delay times 
\footnote {Since $\msec \propto \tage^{-0.3}$ at ages older than 1 Gyr, 
mass distributions as steep as $n(\msec) \propto \msec^{-4.5}$ are required
in order to have a flat \fiasd at late \tage.}.

However, while the distribution function of the
secondary masses of primordial binaries could well be represented by a 
Salpeter IMF, the distribution $n(\msec)$ in 
Eq. (\ref{eq_fiasd}) refers to the secondaries in systems which 
{\em eventually give rise to a SNIa event}, and as
such suffers from some limitations. This is illustrated in the next section.

\subsection{The distribution function of the secondary masses}

The commonly adopted scheme in the population synthesis computations deals with
binary systems in which the primaries follow a
power law distribution with slope $-\alpha$, and the mass ratios 
$q=\msec/\mpri$ follow a power law distribution with a slope $\gamma$. 
Then, the number of binaries with primary masses in ($\mpri;\mpri+d\mpri$)
and secondaries in ($\msec;\msec+d\msec$) is :

\begin{equation}
n(\mpri,\msec)\,\dif \mpri\,\dif \msec = n(\mpri)\,f(q)\,\dif \mpri\,\dif q .
\end{equation}

The distribution function of the secondaries in SNIa progenitor systems, 
is obtained by summing over all 
possible primaries, ranging from a minimum value (\mprii) to 8 \msun: 

\begin{eqnarray}
n(\msec) & \propto & \int_{\mprii}^{8} 
{\msec^{\gamma} \, \mpri^{-(\alpha+\gamma+1)} \, \dif \mpri} \nonumber \\
& \propto & 
\msec^{-\alpha} \cdot
[(\msec/\mprii)^{\alpha+\gamma}-
(\msec/8)^{\alpha+\gamma}] .
\label{eq_phitsc}
\end{eqnarray}

\acap
The minimum mass for the primary (\mprii) is either equal to \msec\ (as in any
binary system), or more massive than
this, if the primary has to produce a sufficiently massive CO WD in order to
lead to a SNIa explosion. Thus:

$$\mprii=\max(\msec,\mprin)$$

\acap
where \mprin\ is the mass of the primary following a specific 
constraint on the minimum mass of the CO WD. The restriction of the integration
to systems with primaries more massive than \mprin\ has important 
consequences on
the distribution function of the delay times, and then, on the SNIa rate 
past a burst of SF, as will be seen in the following.
For Chandrasekhar explosions, \mprin\ is derived by requiring that 

\begin{equation}
\mwdn + \epsilon \cdot \msece = 1.4
\label{eq_mwd}
\end{equation}

\acap
where \mwdn\ is the minimum acceptable mass for the WD, \msece\ is the 
envelope mass of the evolving secondary, and $\epsilon$ is an efficiency 
parameter. 
In principle \msece\ varies with the evolutionary stage at which
the second Roche Lobe Overflow (RLO) occurs, and then depends on the 
separation of the binary 
system. In practice, since the second mass transfer will not necessarily
occur with 100\% efficiency, this detail can be neglected,
to consider a representative relation between \msec\ and \msece. 
For example, the results of case B RLO in Nelemans et al.
(\cite{Nele01}) can be represented by the following relation between
the initial mass (\msec) and its remnant (\msecc):

\begin{equation}
\msecc  =  \max \{ 0.3;0.3+0.1(\msec-2);0.5+0.15(\msec-4)\} .
\label{eq_mwdb}
\end{equation}
 
\acap
By considering $\msece = \msec - \msecc$ it
turns out that smaller secondaries typically have smaller envelopes to
donate to their WD companion. As a consequence, 
the minimum acceptable WD mass \mwdn\ increases as \msec\ decreases,
and so does the mass of its progenitor \mprin.
An additional lower limit on the mass of the primary comes from the
requirement that its remnant should be a CO, rather then a Helium, WD.
Stars less massive than $\sim$ 2 \msun\ develop a degenerate
Helium core, and evolve along the Red Giant Branch up to
the Helium flash, which occurs when the stellar radius is
of a few hundred \rsun\ (see e.g. Fig. 1 in Yungelson \cite{Lev05}). 
Therefore, these stars can provide
CO WDs only if the separation of the primordial binary exceeds
$\sim$ 400 \rsun. Since the
distribution function of the primordial separations \azero\ scales as 
$\azero^{-1}$ (see, e.g. Iben \& Tutukov \cite{IT84}) systems with a primary
less massive than 2 \msun\ are much more likely to produce a Helium
rather than a CO WD. For this reason, the contribution to SNIa from systems 
with \mpri\ smaller than 2 \msun\ is neglected here, and I consider:

\begin{equation}
\mprin = \max \{ 2., 2. + 10.(\mwdn - 0.6) \}
\label{eq_mwdc}
\end{equation}

\acap
where the relation between \mprin\ and its remnant \mwdn\ 
describes the results of case C RLO in Nelemans et al.(\cite{Nele01}),
and represents very well the empirical determination of the 
initial - final mass relation by Williams, Bolte \& Koester 
(\cite{Williams}).  

In summary, as \msec\ decreases, the lower limit of integration in
Eq.(\ref{eq_phitsc}) is first set to \mprii=\msec, down to \msec=2 \msun;
then it stays constant and equal to 2 \msun\ until the minimum WD mass
for a successful SNIa event (\mwdn\ from Eq.(\ref{eq_mwd})) becomes larger
than 0.6 \msun. From that point on, \mprii\ increases with
decreasing \msec. 
At some value of \msec, \mprii\  becomes equal to 8\msun: this marks the
minimum secondary mass suitable for a successful SNIa event.

The scheme adopted in Greggio and Renzini (\cite{IO83}) is slightly 
different from the one just described. It assumes that the total mass of the
primordial binary \mb\ follows a power law with slope $-\alpha$, 
and that the mass ratio $\mu$=\msec/\mb\ is distributed according to 
$f(\mu) \propto \mu^{\gamma}$. In this case:

\begin{eqnarray}
n(\msec) & \propto & \int_{\mbi}^{\mbs} 
{\msec^{\gamma} \, \mb^{-(\alpha+\gamma+1)} \, d\mb} \nonumber \\
& \propto & 
\msec^{-\alpha} \cdot [(2\msec/\mbi)^{\alpha+\gamma}-(2\msec/\mbs)^
{\alpha+\gamma}] 
\label{eq_phitgr}
\end{eqnarray}

\acap
where \mbs = 8+\msec\ and \mbi =\mprii +\msec.

It is worth remarking that in the original formulation by Greggio \& Renzini (\cite{IO83}), 
\mbi\ was required to be
larger than a minimum value (e.g. 3 \msun) irrespectively of the mass of the 
evolving secondary.
This fixed limit does not describe the one to one correspondence between the
mass of the evolving secondary and delay time of the SNIa precursor. 
At given delay time, the amount of mass that can be donated to the 
accreting white dwarf is virtually fixed, which implies a minimum white dwarf
mass, in order to add up to the Chandrasekhar limit.  
A revision of the Greggio \& Renzini (\cite{IO83}) SNIa rate 
has been presented in Greggio (\cite{IO96}).

Equations (\ref{eq_phitsc}) and (\ref{eq_phitgr})
are very similar, but not quite the same. In both cases,
the distribution function of the secondaries follows a power law with slope
$-\alpha$, corrected by a factor which describes the limitations imposed on
the masses of the primaries in order to secure the SNIa explosion.
As \tage\ increases, \msec\ decreases and, as long as \mprii = \msec, the 
correction factor increases toward unity. As soon as \mprii\ becomes greater
than \msec,
the correction factor starts decreasing, due to the loss of systems with
primary mass between \msec\ and \mprii.

\begin{figure}
\resizebox{\hsize}{!}{\includegraphics{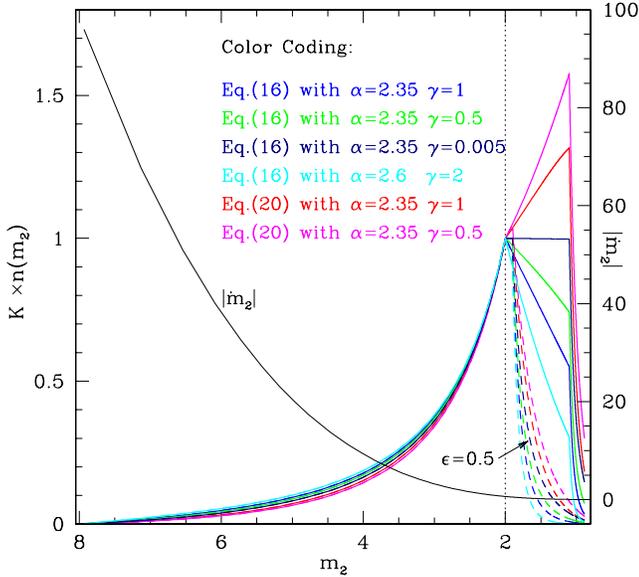}}
\caption{The distribution function of the secondary masses in 
the SD model, for Chandra exploders as obtained with 
Eqs. (\ref{eq_phitsc}) and (\ref{eq_phitgr}) for selected values of 
the $\alpha$ and $\gamma$ 
parameters. Solid lines have been obtained with $\epsilon=1$, dashed lines
with $\epsilon=0.5$ in Eq. (\ref{eq_mwd}). The black curve shows the 
derivative of \msec\ on a linear scale (right axis).}
\label{fig_nm2}
\end{figure}

Figure (\ref{fig_nm2}) shows the distribution 
(in arbitrary units) of the secondaries in systems which lead to 
Chandrasekhar explosions,
under a variety of hypothesis for the power law slopes $\alpha$\ and $\gamma$,
and for the two formulations given by Eqs. (\ref{eq_phitsc}) and 
(\ref{eq_phitgr}), as labeled. 
%

The distribution function of the secondaries in systems which produce a 
SNIa is shaped after the behavior of the minimum mass for the
primary as \msec\ decreases: it
shows a first abrupt change of the slope when \msec\ drops 
below 2 \msun, that is when \mprii\ is set to 
2 \msun\ independently of \msec.
A second abrupt change of the slope appears when
the envelope mass of the secondary is so small
that $\mprii > 2 \msun$ is required in order to
build up to the Chandrasekhar mass. From this point on, the distribution
function of the secondaries steeply decreases as \mprii\
increases.
Smaller values for $\epsilon$ imply an earlier occurrence of this regime,
which in the illustrated case ($\epsilon$=0.5) appears soon after 
\msec\ has gone below 2 \msun.

For the same values of $\alpha$ and $\gamma$, the Greggio \& Renzini 
(\cite{IO83}) formulation (red curves) yields a larger fraction 
of systems with low secondary mass, compared to the formulation generally 
adopted in the population synthesis codes. As a result it leads to 
comparatively larger rates at late epochs.

The distribution $n(\msec)$ also appears very sensitive to the 
$\alpha$ and $\gamma$ parameters. Most noticeably,  
the flatter $\gamma$, the larger the fraction of systems with small
secondaries.  
Figure \ref{fig_nm2} also shows the time
derivative of \msec : 
the combination of the two factors clearly produces an early maximum
for the \fiasd\ function which is given by Eq.(\ref{eq_fiasd}). 

\subsection{The distribution function of the delay times}

\begin{figure}
\resizebox{\hsize}{!}{\includegraphics{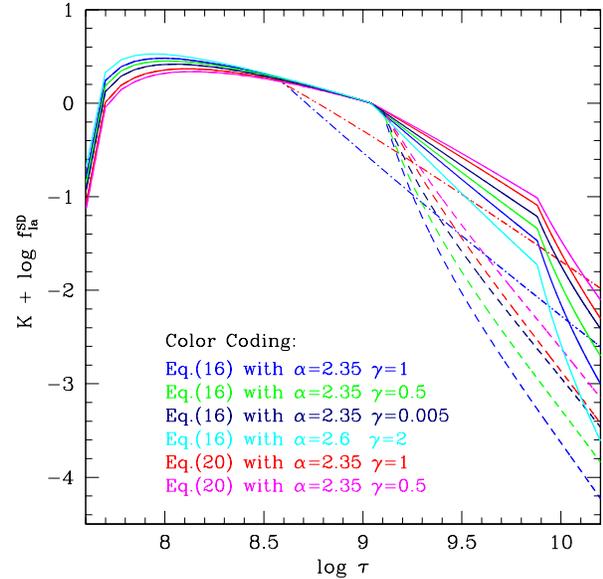}}
\caption{Distribution function of the delay times for 
the SD model for the same choices of the parameters as in
Fig.~\ref{fig_nm2}. In addition, the dot-dashed lines show the
distribution function for the Sub-Chandra models in the case 
($\alpha,\gamma$)=(2.35,1), using
Eq. (\ref{eq_phitsc}) (blue) and Eq. (\ref{eq_phitgr})
(red).} 
\label{fig_fiasd}
\end{figure}

Figure \ref{fig_fiasd} shows the resulting distribution function of 
the delay times, with the same color and line coding as in Fig.
\ref{fig_nm2}. The two abrupt changes of the slope just reflect
those appearing in the $n(\msec)$ function. Figure \ref{fig_fiasd} 
also shows the distribution functions of the delay times expected for the 
Sub--Chandra exploders as dot-dashed lines.
These are obtained from Eqs (\ref{eq_phitsc}) and (\ref{eq_phitgr})
with the following criterion for \mprii: a minimum
WD mass of 0.7 \msun\ is required to secure the explosion,
along the lines suggested by Woosley \& Weaver (\cite{Stan94}) models.
This corresponds to a minimum primary mass of 3 \msun\ (see the second
term of the RHS in Eq. (\ref{eq_mwdc})), so that the first cusp occurs
as early as 0.4 Gyr.
In order to have a Sub--Chandra SNIa, a Helium
layer of about 0.15 \msun\ needs to be accumulated on top of the CO WD,
which corresponds to $(\epsilon \cdot \msece) \ge 0.15 \msun$.
This is a limit on \msec\ which implies that the single burst SNIa rate for 
this model drops to zero at a delay time equal to the evolutionary lifetime
of such \msec. 
Since the evolutionary mass at 15 Gyr is $\sim$ 0.9 \msun, this limit is
inactive for the whole Hubble time, provided that $\epsilon \ga 0.3$, 
as is considered for Fig. \ref{fig_fiasd}. 

The distribution function of the delay times for the SD model 
is basically shaped according to the limits on the mass of the primaries
in systems which eventually produce a SNIa. Four regimes can be identified:

\acap
1) Regime A, in which \fiasd\ shows an initial steep rise followed by a 
mild decrease. In this regime \mprii=\msec, implying that at longer
delay times there is an increasing range of \mpri\ which contribute
to the SNIa explosions. The (wide) maximum results from 
the interplay between $n(\msec)$ and $|\dot{\msec}|$, the latter being a 
decreasing function of \tage.

\acap
2) Regime B, which sets in when \msec\ falls below the minimum primary
mass suitable for the explosion. 
Such minimum mass is chosen here equal to
2 and 3 \msun\ respectively for Chandra and Sub--Chandra exploders.
In Regime B, \mprii\ is kept constant and equal to this minimum primary 
mass; the distribution function of the delay times decreases with
a steeper slope compared to Regime A, due to the loss of systems with
$\msec \le \mpri \le 2$.
It can be shown that in this regime, the \fiasd\ function is mostly sensitive
to the parameter $\gamma$, according to 
$\log \fiasd \propto \tage^{-1.44-0.3\gamma}$.

\acap
3) Regime C, starting when primaries more massive than 2 \msun\ are required
in order to build up to the Chandrasekhar mass. In Regime C, \mprii\
increases from 2 to 8 \msun\ as \tage\ increases; consequently
\fiasd\ shows a steep decline. For Sub--Chandra exploders
Regime C does not exist.

\acap
4) Regime D, in which \fiasd=0: this sets in when \mprii=8 \msun\ is
required to produce a Chandrasekhar explosion, or when 
the envelope mass of \msec\ drops below the $(0.15/\epsilon)$ limit.
Regime D is not shown in Fig.~\ref{fig_fiasd}.
The value of \tage\ at which this regime sets in defines \taux\ for the
SD model.

For the Chandra exploders, the delay time at which Regime C sets in varies
with the efficiency of the accretion process. The response of the CO WD to
accretion crucially depends on the accretion rate, as illustrated
in Hachisu \& Kato (\cite{HK01}). Only if the accretion rate is in a small
range around $10^{-7}$ \msun/yr does steady burning occur on top of the WD,
so that the mass of the donor is efficiently used to increase the mass of
the WD. For both lower and higher accretion rates, part of the donated mass
is lost, either following  H--shell flashes (novae explosion), for low
accretion rates, or in a thick wind, for high accretion rates.
Therefore, the accretion efficiency, i.e. the ratio between the accreted
and the donated matter, is unlikely to be close to unity, as adopted for
the solid lines in Fig.~\ref{fig_fiasd}. 
The lower the accretion efficiency, the earlier regime C sets in,
and regime B is suppressed when $\epsilon \lesssim 0.5$, as shown by
the dashed lines in  Fig.~\ref{fig_fiasd}.


The shape of the \fiasd\ function derived here is quite general, 
since it just reflects the product of the time derivative of the 
evolutionary secondary mass, and the 
distribution $n(\msec)$.

\section {Double Degenerates}

In the DD model, the first part of the close binary evolution is the
same as in the SD model, but, following the expansion of the secondary
component, a Common Envelope phase (CE) sets in, eventually leading to
the complete ejection of the CE itself. In this scenario, accretion on top of
the WD, if any, is neglected. The system emerges from the CE as a close 
binary composed of two WDs, which are bound to eventually merge due to the 
emission of gravitational wave radiation. If the total mass of the DD 
system exceeds the 
Chandrasekhar mass, explosion occurs as a SNIa event. The delay time in 
this case is

\[\tage = \taun + \taugr\]

\acap 
where \taun\ is the nuclear lifetime of the secondary, and \taugr\ is the 
gravitational delay (Landau \& Lifshitz \cite{Landau}): 

\begin{equation}
\taugr = \frac {0.15\,\,\aff^4}{(\mpwd+\mswd)\mpwd\mswd} \, \mathrm{Gyr}
\label{eq_taugr}
\end{equation}

\acap
\aff, \mpwd\ and \mswd\ being respectively the separation and component 
masses of the DD system, in solar units.


To derive the distribution function of the delay times one needs
to map the distribution of
the primordial systems in the space (\azero ,\,\mpri ,\,\msec) into that
of the final systems in the space (\aff ,\,\mpwd ,\,\mswd).
Rather then performing Montecarlo simulations I consider here some
general aspects, with the aim of characterizing the shape of the 
\fia\ function.
 
I restrict to systems with $2 \le \mpri,\msec \le 8$, from which most
double CO WDs form, as argued in Sect. 3.1. 
Typically, the WD mass of both components ranges 
between 0.6 and 1.2 \msun\ (see e.g. Eq. (\ref{eq_mwdc})), 
so that \taun\ ranges between 0.04 and 1 Gyr. 
The gravitational delay 
spans a large range, depending on the final separation: e.g., 
for a (0.7+0.7) \msun\ DD, \taugr\ increases from 0.014 to 18 Gyr when
\aff\ goes  from 0.5 to 3 \rsun. 
The distribution function of the delay times will depend on the distributions
of both \taun\ and \taugr, at least up to total delays of $\sim$ few Gyr, 
with early explosions provided by systems with short \taun\ {\em and} \taugr.
Since the maximum nuclear delay does not exceed $\sim$ 1 Gyr, at  
long \tage\, the SNIa events will come from systems with long gravitational
delays, that is DDs with wide separations and low masses. 

\subsection {The interplay of \taun\ and \taugr}

The following useful approximation for the gravitational delay is justified in 
Appendix ~\ref{appe_fmr}:

\begin{equation}
\taugr = 0.6\frac{\aff^4}{\mdd^3} \, \mathrm{Gyr}
\label{eq_taugrap}
\end{equation}

\acap
where \mdd\ is the total mass of the DD system (again in solar units).
This formula results from considering the restrictions on the masses
of the two WD components. 
In this approximation the gravitational delay does not explicitly
depend on \mswd. 
Since \taun\ only depends on \msec, which is in tight correspondence with
\mswd, the relation between the two timescales is very weak:
basically, primordial systems with secondary mass \msec\ will evolve into
a family of SNIa precursors spanning a wide range in separations and total
binary mass, that imply a wide range of gravitational delays.

The integrated distribution of the total delay times can now be constructed:
the contribution (d$n_{\rm n}$) of binaries with given \taun\ to the number of 
systems with total delay time shorter than \tage\ is proportional to
the fraction of them which have \taugr\ shorter than ($\tage-\taun$). 
Indicating this quantity with $g(\tage,\taun)$: 

\begin{equation}
\dif n_{\rm n} =  n(\taun) \cdot g(\tage,\taun)\, \dif \taun 
\label{eq_ntautm}
\end{equation}

\acap
where $n(\taun)$ is the distribution function
of the nuclear timescales, proportional to the \fiasd\ function described in
Sect. 3. The total number of systems with total delay time shorter than 
\tage\ is obtained by integrating the d$n_{\rm n}$ contributions over the 
relevant \taun\ range: 
 
\begin{equation}
\int_{0}^{\tage}n(\tage)\dif \tage =
\int_{\tauni}^{\min(\taunx,\tage)}n(\taun)\,g(\tage,\taun)\,\dif \taun 
\label{eq_ntau}
\end{equation}

\acap
where \tauni\ and \taunx\ bracket the range of nuclear timescales of the
systems which end up in a successful SNIa event.



  

Let's now indicate with $n(\taugr)$ the distribution function of the
gravitational delays of systems with a nuclear delay \taun\ 
(i.e. progeny of systems with a
secondary mass \msec\ whose MS lifetime is equal to \taun), and let 
\taugi\ and \taugx\ be the minimum and maximum 
gravitational delays of such systems.
Formally, the fraction of systems which, having a nuclear delay equal to  
\taun, have also a total delay shorter than \tage\ varies with \tage\ 
according to:



\begin{equation}
g(\tage,\taun) = \left\{
\begin{array}{lll}
0 & \mbox{for \tage $\leq \tage_1$}\\
\int_{\taugi}^{\min(\taugx,\tage-\taun)}n(\taugr)\,\dif \taugr & \mbox{for
$\tage_1 \leq$ \tage $\leq \tage_2$} \\
1 & \mbox{for \tage $\geq \tage_2$}
\label{eq_gttn0}
\end{array}
\right.
\end{equation}

\acap
with $\tage_1 = \taun+\taugi$ and $\tage_2 = \taun+\taugx$.
This equation merely expresses that (i) systems with nuclear delay 
\taun\ have a total
delay ranging between \taun+\taugi\ and \taun+\taugx; 
(ii) the total delay of such systems scales according to the 
distribution function of their gravitational delays.


The shape of $n(\taugr)$
will reflect the distribution of the final separations and of the DD masses. 
The interesting range of gravitational 
delays, which goes from $\sim 0.01$ Gyr (i.e. on the order of \tauni)  
to over the Hubble time, is populated by low mass systems (\mdd=1.4\msun) 
with $0.5 \lesssim \aff/\rsun \lesssim 2.8$, and by high mass systems 
(\mdd=2.4\msun) with $0.7 \lesssim \aff/\rsun \lesssim 4.2$. 
In the next section the outcome of the close binary evolution is examined 
in connection to the possibility
of producing final separations in the reference range
$0.5 \lesssim \aff/\rsun \lesssim 4.5$, which corresponds to the relevant
range of gravitational delays, for the full mass range of the DD systems. 
 


\subsection{Shrinkage during the mass transfer phases}

The mass transfer phase in a close binary system may be
dynamically stable or unstable: in the first case the outcome
is a wide system (occasionally wider than the primordial separation), 
and the secondary may have accreted some of the donor's envelope mass.
In the second case, a CE occurs and (generally) the system shrinks.
The occurrence of one evolutionary channel rather than another depends
on the configuration of the initial binary (e.g. the mass ratio at RLO, 
whether the envelope of the donor is radiative or convective, and so on).
Anyway, the interesting systems are those which suffer a 
substantial degree of shrinkage during their evolution:  
the initial separations of the double CO WD progenitors range roughly 
from 100 to 1000 \rsun, so that the first RLO takes place at all (upper limit),
and that it does so after core Helium exhaustion (lower limit), 
when the star is on the Asymptotic
Giant Branch. Thus, in order to merge within a Hubble time, 
the binary evolution should produce a total shrinkage 
on the order of $\sim$ $10^{-2},10^{-3}$. This can be accomplished through one 
or more CE phases.

The standard CE recipe (e.g. Webbink 1984) relates the initial and final
values of the binary parameters by requiring that the variation of the 
orbital energy is proportional to the binding energy of the envelope of the
donor :

\begin{equation}
\frac{\mdi(\mdi-\mdf)}{R}=\ace \left[ \frac{\mdf\,m}{2\af}-
\frac{\mdi\,m}{2\ai} \right]
\label{eq_ce} 
\end{equation}
 
\acap
where \mdi, \mdf\ are the mass of the donor respectively
before and after the CE, $m$ is the mass of the companion, \ai\ and
\af\ are the separations before and after the CE, $R$ is the radius
of the donor at contact (i.e. the Roche Lobe radius).
\ace\ is parameter roughly describing the efficiency with which the orbital
energy of the binary is used to expel the CE  
\footnote {Different formulations are found in the
literature for Eq.~(\ref{eq_ce}), and correspondingly different meanings for 
\ace. I adopt here the formulation in Nelemans et al. (\cite{Nele01}), with 
the geometrical parameter $\lambda$=1.}: when
smaller than unity the process is very inefficient, and the system emerges
from the CE with a small separation; when $\ace > 1$ other energy 
sources, besides the orbital one, are used to expel the CE.
Typical values considered in the literature, and supported by
hydrodynamical computations, range from \ace=0.5\ to \ace=2 
(Rasio \& Livio \cite{Rasio}).

\begin{figure*}
\centering
\includegraphics[width=17cm]{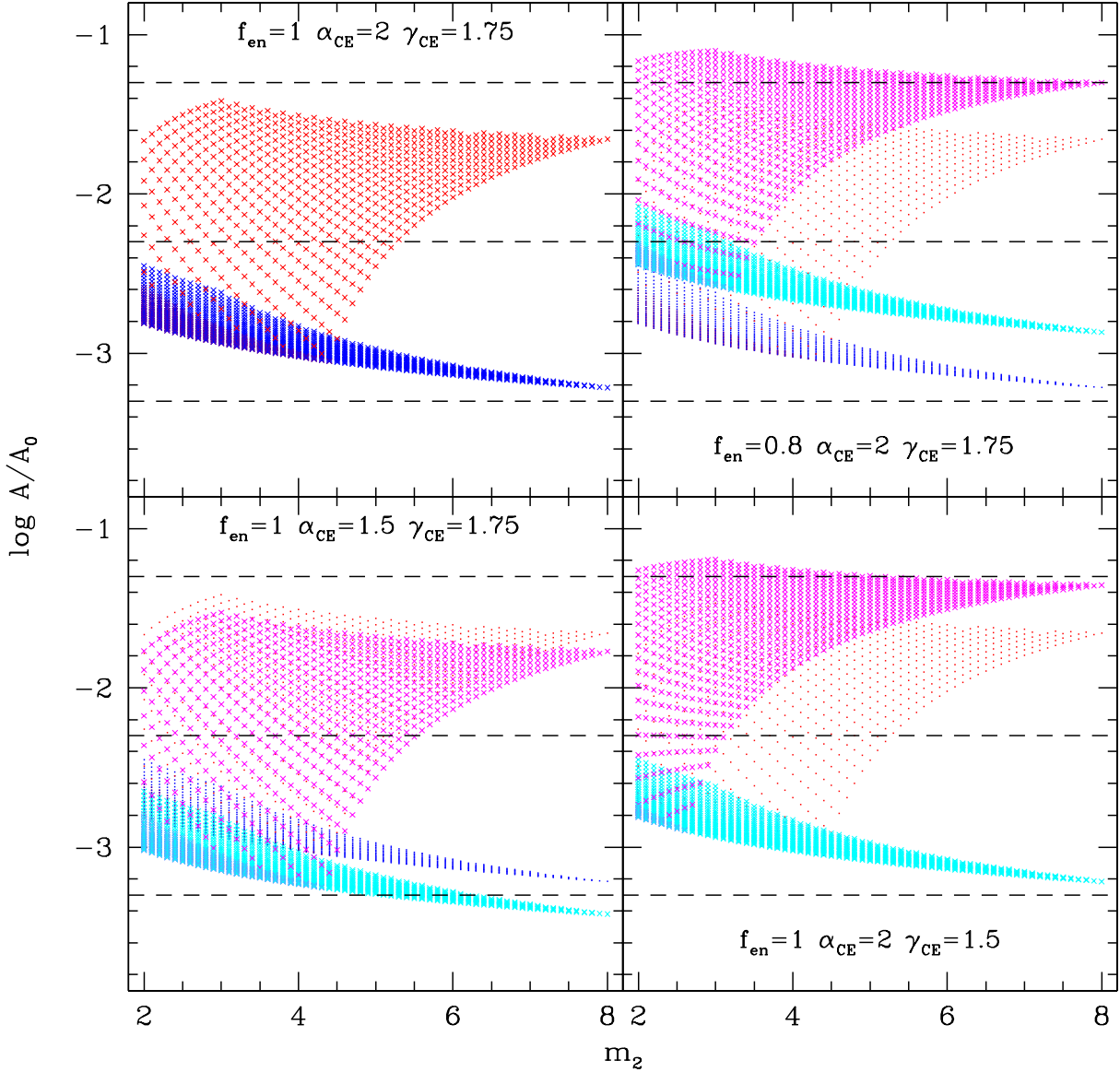}
\caption{Exploration of the outcome of two mass transfer episodes in binaries
with intermediate mass. The ratio between the separations of the final DD
and the primordial systems is plotted as a function of the initial mass of the 
secondary for systems with $\mpri \geq \msec$.
Eq. (\ref{eq_mwdc}), and the Eggleton (\cite{Eggleton}) formula 
for the Roche Lobe radius have been used. Blue and cyan points 
result from the application of Eq. (\ref{eq_ce}) to both mass transfers; 
red and magenta points 
from applying either Eq. (\ref{eq_ne})
or Eq. (\ref{eq_ce}) for the first mass transfer, depending on which of the 
two yields the wider separation \af. 
At given \msec, the larger \mpri\ the smaller \affai\ is. 
Each panel is labeled with the adopted values for
the parameters in Eqs. (\ref{eq_ce}) and (\ref{eq_ne}). The results in
the top-left panel are reproduced in the other panels as small dots
for an easy comparison.}
\label{fig_shrink}
\end{figure*}

However, this formulation fails to explain the observed binary parameters of 
3 double He WDs, and 
Nelemans et al. (\cite{Nele00}) propose an alternative equation for the first
mass transfer, which consists in a parametrization of the system's angular 
momentum loss:

\begin{equation}
\frac{J_{\rm i}-J_{\rm f}}{J_{\rm i}}=\gammace \frac{\Delta M}{M_{\rm B}}
\label{eq_ne}
\end{equation}

\acap
where $J_{\rm i},J_{\rm f}$ are the initial and final angular momenta,
$\Delta M$ is the mass lost from the system (roughly equal to the 
donor's envelope mass), $M_{\rm B}$ is the total mass of the binary
before RLO, and $\gammace$ is a parameter, for which Nelemans et al. 
(\cite{Nele00}) find a value of $\simeq 1.5$
by fitting the data. This formulation has been recently shown to describe
the properties of a larger sample of binary WDs 
(Nelemans \& Tout \cite{Nele05}).

Eq.(\ref{eq_ce}) leads to a dramatic shrinkage, the typical ratio
between the separations after and before the CE phase ($\afai$) being 
on the order of a few $10^{-2}$; instead, for systems with a high mass
ratio $q$ ($=\msec/\mpri$), Eq. (\ref{eq_ne}) leads to modest shrinkage 
($\afai \sim 1$). At low $q$, Eqs. (\ref{eq_ce}) and (\ref{eq_ne}) 
yield similar values, and actually, Eq. (\ref{eq_ne}) is only applicable at
relatively large $q$ (see Appendix ~\ref{appe_shrink}). 
As for the second mass transfer episode, Eq. (\ref{eq_ce}) is generally 
adopted in the literature. I have then explored the results of two different 
evolutionary schemes: 
(i) the two successive RLO are regulated by Eq. (\ref{eq_ce}); (ii)
the first mass transfer occurs according to Eq. (\ref{eq_ne}) (when applicable,
see Appendix ~\ref{appe_shrink} for a detailed description), while the 
second mass transfer is regulated by Eq. (\ref{eq_ce}).
The evolutionary scheme (i) corresponds to the prescriptions used by
e.g. Tutukov \& Yungelson (\cite{Tuyu}), Ruiz-Lapuente \& Canal 
(\cite{Pilar95}), Han (\cite{Han95}), Han et al.
(\cite{Han98}), Yungelson \& Livio (\cite{YL}); 
the evolutionary scheme (ii) corresponds to the prescriptions used
in Nelemans et al. (\cite{Nele01}).
 
The total shrinkage, that is the ratio (\affai) between the separation of the
newborn DD system and the original separation of the binary, is shown in 
Fig.~\ref{fig_shrink} as a function of the mass of the secondary component
in the primordial system: blue and cyan dots refer to the (i), 
red and magenta dots to the (ii) evolutionary schemes. At fixed \msec\ the
points show the effect of varying \mpri: the more massive \mpri\, the smaller 
the \affai\ ratio. The four panels show the effect of varying the parameters
\ace\ and $\gammace$ in Eqs. (\ref{eq_ce}) and (\ref{eq_ne}).
In order to gauge the effect of mass loss through stellar wind, which
may occur prior to the CE phases, 
an additional parameter (\fen) has been considered, 
such that the mass of the donor at each RLO is equal to a fraction 
\fen\ of its initial mass. The sensitivity of the final shrinkage to this
parameter proves significant. 

The two options lead to
vastly different situations: for the scheme (i) the \affai\ ratio appears
confined in a narrow range, around a mean value which depends on \ace\ and
\fen.
In addition, there is a clear trend of \affai\ decreasing as \msec\ increases.
Indeed, more energy is required to expel the more massive CE in the more 
massive binaries. At fixed \ace\ this implies that the systems shrink more
and end up with a smaller \affai\ ratios. 
The (ii) prescriptions, instead, produce a wide range of \affai\ ratios at 
almost every \msec. This means that systems with the same \msec\ and \azero\
can end up very wide or very close depending on the mass of the companion.

Turning now to consider the quantitative value of \affai,
the three lines in Fig. ~\ref{fig_shrink} show the
levels $\affai=5\cdot10^{-4},5\cdot10^{-3},5\cdot10^{-2}$.
Recalling that the initial separations of the double CO WD progenitors 
range roughly from 100 to 1000 \rsun, such levels correspond respectively to  
final separations of $\simeq 0.05, 0.5$ and 5 \rsun\ for the initially 
closest systems;
to $\aff \simeq 0.5, 5$ and 50 \rsun\ for the initially widest ones.
Inspection of Fig.~\ref{fig_shrink} shows that the evolutionary
scheme (ii) is capable of producing final separations in a very wide 
range, well including the range $0.5\rsun \la \aff \la 5\rsun$, leading
to merging within a Hubble time.
On the contrary, the scheme (i) appears to produce very small 
\affai\ ratios, so that
only the initially widest binaries manage to merge on timescales on the order
of some Gyr. In addition, the correlation between \affai\ and \msec\ 
implies that lower mass systems have longer gravitational delays, as 
will be better illustrated in the next section. 

It's important to notice the high sensitivity of the \affai\ ratio
on the parameters (\ace,\fen,$\gammace$); this, coupled with the high
sensitivity of \taugr\ on \aff\ suggests that the results of the binary
evolution from the population synthesis codes are very dependent 
of the exact recipe used.
At the same time, the correspondence between the (\azero,\,\mpri,\,\msec) and
(\aff,\,\mpwd,\,\mswd) is likely to be rather loose, and even more so if
a distribution of (\ace,\,\fen,\,$\gammace$) values is realized in nature. 
In this respect,
it is worth recalling that the computations here adopt a unique 
initial-final mass relation, while in reality,
at a given initial mass, the remnant mass spans a (small) range, in relation
to the precise point in the evolution at which the mass transfer takes place.
Therefore, for each pair (\mpri, \msec) there will be a distribution of 
\affai, around the corresponding point in Fig. ~\ref{fig_shrink}.

This exploration of the results of the CE phases suggests to consider 
the two following extreme characterizations for the $n(\taugr)$ function:  

\begin{itemize}

\item either, irrespectively of \msec, the close binary evolution produces
a wide distribution of \aff\ and \mdd, and these two variables are 
virtually independent;

\item or the close binary evolution leads to a narrow distribution of the 
\affai\ ratio, so that the initially closest binaries merge in a short time, 
and the initially widest binaries tend to populate the long \taugr\ tail of the
distribution. In addition, the most massive binaries tend to end up with
the smallest final separation, hence merge more quickly.

\end{itemize}

\acap
The first characterization is more appropriate for the scheme (ii)
of evolution, and will be referred to as {\em WIDE DD}s;
the second, suggested by the results of the evolutionary scheme (i),
will be referred to as {\em CLOSE DD}s. Since both schemes provide
very small \affai\ ratios for some values of the parameters of the primordial
binary system, the minimum gravitational delay \taugi\ is likely to
be very short. The dependence of the distribution function of the delay
times on the parameter \taugi\ will be explored later, and only for
the {\em WIDE DD} scheme. The maximum gravitational delay \taugx\ is instead
an important parameter for the characterization of the two schemes, as 
shown in the next sections.

\begin{figure}
\resizebox{\hsize}{!}{\includegraphics{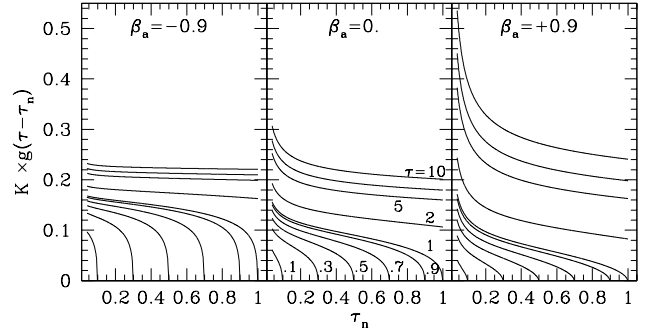}}
\caption{The $g(\tage,\taun)$ function for {\em WIDE DDs},
with \fniii=\fni, for three values of 
the $\betaa$ parameter. The functions, plotted on arbitrary units, are shown
for \tage = 0.1, 0.3, 0.5, 0.7, 0.9, 1, 2, 5, 7 and 10 Gyr. 
A very small minimum gravitational delay \taugi=0.001 Gyr has been
used.} 
\label{fig_gfun1}
\end{figure}

\subsection {The distribution $g(\tage,\taun)$}

The fraction of 
systems which, having a nuclear timescale \taun\, manage to
merge within a timescale shorter than $\taugr=\tage-\taun$ depends on the 
distribution of the gravitational delays.
In this section the expressions for the $g(\tage, \taun)$ function 
are derived for the {\em WIDE DD} and {\em CLOSE DD} schemes , based on the
two different characterizations.

\subsubsection {WIDE DDs}

For the {\em WIDE DD} scheme the assumptions are:

\begin{itemize}
\item \mdd\ and \aff\ are independent variables;
\item the minimum gravitational delay \taugi\ is independent of \msec;
\item the maximum gravitational delay \taugx\ is larger than $\tage-\taun$ 

for all \taun\ (at least for a total delay time up to the Hubble time);
\item the distribution function of the final separations follows a power law:
$n(A) \propto A^{\betaa}$.
\end{itemize}

Under these conditions, in Appendix ~\ref{appe_gnele} the following  formula 
is derived:


\begin{equation}
g(\tage,\taun) \propto \left\{
\begin{array}{ll}
0 & \mbox{for \tage$\leq \tage_1$} \\
\fniii \,\left[ (\tage-\taun)^{\betaat}-\taugi^{\betaat}
\right]
& \mbox{for $\tage_1 \leq$\tage$\leq \tage_2$}
\label{eq_gttn}
\end{array}
\right.
\end{equation}

\acap
where $\betaat = 0.25(1+\betaa)$, $\tage_1 = \taun+\taugi$ , 
$\tage_2 = \taun+\taugx$ and 
\fniii\ is a function of \taun, for which two cases are considered:

\begin{equation}
\fniii = \left\{
\begin{array}{ll}
\fni = & \mddt^{0.75+0.75\betaa} \\
\fnii = & \mddx^{1.75+0.75\betaa} - \mddn^{1.75+0.75\betaa} 
\label{eq_fniii}
\end{array}
\right.
\end{equation}

\acap
with

\begin{displaymath}
\mddn=\max (1.4,\mswd+0.6); \, \, \, \, \, \mddx=\mswd+1.2;
\end{displaymath} 
\begin{displaymath}
\mddt=1.4+(\msec-2)/6
\end{displaymath}

\acap
and \mswd\ given by Eq. (\ref{eq_mwdc}).
The first case in Eq. (\ref{eq_fniii}) applies when there is a tight
correlation between \msec\ and its \mdd\ progeny; the second applies
when a wide distribution of \mdd\ is obtained from systems with the same
\msec.

Fig.~\ref{fig_gfun1} illustrates the $g(\tage,\taun)$ function (computed
with the \fni\ factor).
In general, as \tage\ increases, the $g$ function increases at 
every \taun, since at longer delays more systems fulfill the condition
$\taugr+\taun \leq \tage$ at each \taun. 
At the same time, for increasing \taun\ the number of systems with
total delay smaller than a fixed \tage\ decreases, partly because of the 
decrease 
of the range of gravitational delays which fulfill the condition 
$\taugr \leq \tage-\taun$. A more thorough explanation of the trend of the
$g$ function can be found in Appendix ~\ref{appe_gnele}.
For $\betaa=-0.9$ most DDs are born with small separations, and the 
dependences on
both \taun\ and \tage\ are milder, since both the above mentioned effects are 
less relevant. Conversely, when $\betaa=+0.9$ the distribution of the 
gravitational delays is skewed toward the large \taugr\ values: the
fraction of systems with total delay up to \tage \ greatly increases with
\tage, as systems with longer \taugr\ are included. 

It is worth pointing out that for $\betaa < -1$ the $g$ function is very
sensitive to \taugi. This regime 
corresponds to distributions $n(\aff)$ highly peaked at the minimum 
separation, a possibility that likely provides exceedingly small SNIa rates at 
late epochs in early type galaxies.

\subsubsection {CLOSE DDs}

The distinctive features of the end product 
of the {\em CLOSE DD} scheme of evolution
appear to be that (i) the average \affai\ ratio is very small; (ii) it
is correlated with the mass of the secondary in the primordial system.
As a result, the gravitational delays can be very small even for the systems
with maximum primordial separation, especially for the most massive DDs.

\begin{figure}
\resizebox{\hsize}{!}{\includegraphics{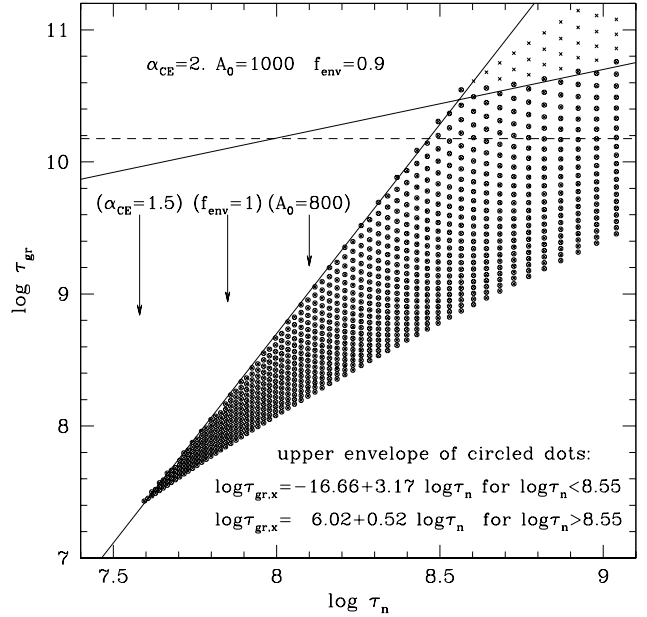}}
\caption{Exploration of the gravitational delay of
DDs from systems with a primordial separation of \azero=1000 \rsun, 
evolved through two CE phases regulated by Eq. (\ref{eq_ce}),
as a function of the MS lifetime of the secondary. Both ages are expressed
in years. Circled dots mark systems with
$\mdd \geq 1.4\msun$. The dashed line shows the 15 Gyr level;
the arrows show the average shift which applies when, leaving the other 
parameters unchanged, one adopts in turn different values for \fen, \ace\
or \azero\ as specified.} 
\label{fig_taugx}
\end{figure}

Fig.~\ref{fig_taugx}  shows the gravitational delay in this evolutionary
scheme for systems born with \azero=1000 \rsun, and having adopted 
\ace=2 and \fen=0.9. Notice that these are about the widest interacting 
binaries: the components of systems born with a larger \azero\ evolve
as single stars, and do not provide SNIa, at least in a Hubble time.
In Fig.~\ref{fig_taugx}, for every \msec\ (i.e. \taun), \mpri\ is 
decreased from 8 to \msec\footnote {Unlike for the {\em WIDE DD} scheme, 
it seems
adequate to restrict here to systems with $\mpwd \geq \mswd$.}, leading to 
longer gravitational delays (\mdd=\mpwd+\mswd\ decreases). 
The circled dots refer to systems with $\mdd \geq 1.4$, and therefore
suited to successful SNIa events. The upper envelope of the gravitational
delays of the SNIa precursors is well represented by:

\begin{equation}
\log \taugx = \min (-16.66 + 3.17\,\log \taun ; 6.02 + 0.52\,\log \taun)
\label{eq_taugxc}
\end{equation}

\acap
with the delay times expressed in years.
By varying the parameters \ace\ and $f_{\rm en}$ 
the linear regressions shift vertically, approximately maintaining 
their slopes. The vertical arrows drawn in Fig.~\ref{fig_taugx} show the 
amount of this shift for the indicated value of the parameters.

The locus $\taugr= 15$ Gyr is shown as a dashed line in Fig.~\ref{fig_taugx}:
it appears that the possibility of realizing gravitational delays as long as
the Hubble time is related to a fine tuning of the involved parameters,
so that the shrinkage due to the two CE phases is not too severe.
The systematic increase of the maximum gravitational delay with increasing
nuclear lifetime of the secondary reflects the smaller shrink of systems
with smaller \msec, which is related to the minor amount of energy
required to expel a less massive CE. For the same reason, at given \msec, 
a less massive primary implies a smaller amount of shrinkage at the first CE.  
%
Therefore, for the {\em CLOSE DD} scheme it seems appropriate to adopt a 
parametrization which emphasizes the systematic depletion of systems with 
long \taugr\ as \taun\ decreases, or the increase of the maximum gravitational
delay in systems with smaller secondary mass.
For simplicity, I consider directly \taugr\ as the independent variable,
and assume that for each \msec\ the
differential distribution of the gravitational delays scales as 
$n(\taugr)\propto\taugr^{\betag}$ between a minimum (\taugi) and a maximum 
value (\taugx). Admittedly, this choice is rather arbitrary; I just notice 
that, if the distribution of \taugr\ is mainly controlled by the distribution
of \aff, one can write:

\[n(\taugr) \dif \taugr \propto n(\aff) \dif \aff\]

\acap
which, in combination with the relations $\taugr \propto \aff^{4}$ and
$n(\aff) \propto \aff^{\betaa}$ becomes $n(\taugr)\propto\taugr^{\betag}$
with $\betag = -0.75+0.25\betaa$. 

With this assumption, the fraction of systems which manage to merge within
(\tage-\taun) is:


\begin{equation}
g(\tage,\taun) = \left\{
\begin{array}{lll}
0 & \mbox{for $\tage-\taun \leq \taugi$} \\
\frac{(\tage-\taun)^{1+\betag}-\taugi^{1+\betag}}{\taugx^{1+\betag}-\taugi^{1+\betag}} & \mbox{for $\taugi \leq \tage-\taun \leq \taugx$}\\ 
1 & \mbox{for $\tage-\taun \geq \taugx(\taun)$}
\end{array}
\right.
\label{eq_gttc} 
\end{equation}

\acap
where \taugx\ is an increasing function of \taun.
It is further assumed that the minimum gravitational delay \taugi\ is 
independent of \msec\ and 
small (compared to \tauni), while the maximum gravitational delay is correlated
to \taun\ via Eq. (\ref{eq_taugxc}).

\begin{figure}
\resizebox{\hsize}{!}{\includegraphics{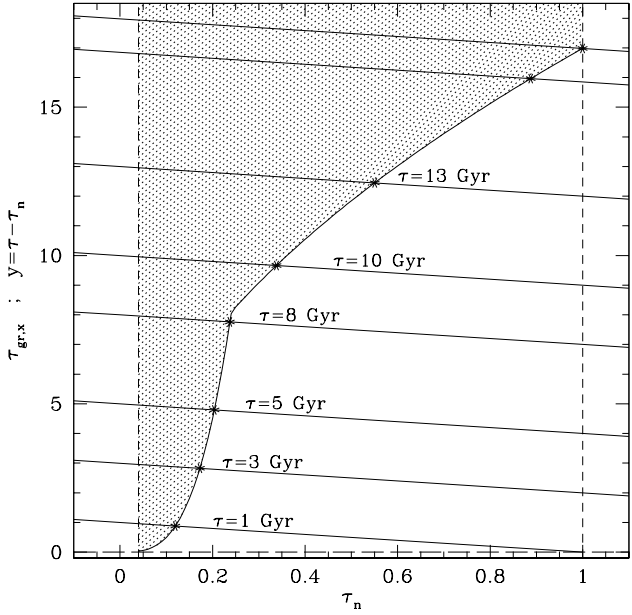}}
\caption{Illustration of how the \taust\ point, defining the branching 
in the {\em CLOSE DD} distribution of gravitational
delays, shifts with increasing total delay \tage. The straight lines show
the loci  $y=\tage-\taun$ for a few values of \tage; the curve shows
the locus \taugx(\taun): for illustrative purposes, the 
relation $\log\taugx=\min[-16.66+3.17\log\taun;5.4+0.52\log\taun]$ is used. 
Delay times are in Gyr. The asterisk marks the branching
occurring at \taun=\taust. The shaded region shows the domain over which
g(\tage,\taun)=1. The dashed lines indicate the total range of nuclear
delays, corresponding to the range of \msec\ considered here, i.e. from
8 to 2 \msun.} 
\label{fig_taunst}
\end{figure}

To better understand the behavior of the $g$ function, Fig.~\ref{fig_taunst}
illustrates how the second branching in Eq. (\ref{eq_gttc}) 
varies as \taun\ increases.
The parallel straight lines show the 
$y=\tage-\taun$ loci for selected values of \tage, ranging from 1 to 
$\sim$ 18 Gyr; the curve is one example for the \taugx\ locus 
(different from Eq. (\ref{eq_taugxc}) for illustrative purposes). 
The intercept between each straight line and the \taugx\ curve 
(in \taun=\taust) defines the branching of Eq.(\ref{eq_gttc}): 
at given \tage, systems with \taun\ shorter than \taust\ have all 
merged, while only a fraction of the systems with $\taun > \taust$ have
already exploded \footnote{Notice that for $\tage\leq\tauni+\taugx(\tauni)$ 
there's no intersection between the two loci: in this range, the total
delay considered is so short that $g(\tage,\taun) < 1$ even for
the shortest \taun. This regime is not visible in Fig.~\ref{fig_taunst} 
because, for the chosen relation, $\tauni+\taugx(\tauni)=$0.07 Gyr.}.
Thus, over the shaded portion of the plane, $g(\tage,\taun)=1$, i.e.
the fraction of systems which, having a nuclear delay \taun, merge within
a total delay \tage\ is constant and equal to 1.
Clearly, {\em at} long total delays, the area in the parameter space over 
which systems have already merged widens, and only those with long \taun,
and therefore long \taugr, still give a variable contribution.
When $\tage \geq \taunx+\taugx(\taunx)$, $g(\tage,\taun)=1$ for
whatever \taun, that is for all the SNIa precursors.
Thus, in this scheme, the maximum delay time of the whole DD population from
an instantaneous burst of SF is \taux = \taunx+\taugx(\taunx).
 

 
By varying the parameters of the close binary evolution (e.g. \ace, \fen) 
the \taugx(\taun) locus shifts.
If evolution leads to systems which are all very close, the maximum 
gravitational delay will be small, and soon \taux\ will be reached, at
which point the distribution function of the delay times for the
{\em CLOSE DD}s drops to 0.

\subsection{The Distribution Function of the Delay Times for DDs}

The distribution function of the delay
times for the DDs is obtained by computing the derivative of 
Eq.(\ref{eq_ntau}):


\begin{equation}
\fiadd(\tage) =
\frac{\dif}{\dif\tage} \, \int_{\tauni}^{\min(\taunx,\tage)}n(\taun)\,g(\tage,\taun)\,\dif \taun .
\label{eq_fiadd0}
\end{equation}

\acap
where $n(\taun)$ is the distribution function of the nuclear delays of
the SNIa progenitors,
that is the \fiasd\ function derived in Sect. 3.2.





Eq.(\ref{eq_fiadd0}) is solved by applying the Leibniz integral rule, 
as shown in Appendix ~\ref{appe_leib}. The final function is:

\begin{eqnarray}
\fiadd(\tage) & \propto &
\int_{\tauni}^{\min(\taunx,\tage)}{n(\taun)\,\stepw(\tage,\taun)
           \,\dif \taun} \,\,
                \mbox{for {\em WIDE DD}s} \label{eq_fiaddw}\\
\fiadd(\tage) & \propto &
\int_{\tinf}^{\min(\taunx,\tage)}
       {n(\taun)\,\stepc(\tage,\taun)
          \,\dif \taun} \,\,
                 \mbox{for {\em CLOSE DD}s} \label{eq_fiaddc}
\end{eqnarray}


\acap
complemented with 

\begin{equation}
      \fiadd(\tage) = 0 \hspace{1cm} \mathrm{for} \,\, \tage \leq \taui\,\, 
\mathrm{and\,\, for} \,\,\, \tage \geq \taux 
\end{equation}

\acap
where

\begin{equation}
\stepw(\tage,\taun) = \left\{
\begin{array}{ll}
\fniii (\tage-\taun)^{-0.75+0.25\betaa} & 
\mbox{for $\taun \leq \tage-\taugi$} \\
0 & \mbox{for $\taun \geq \tage-\taugi$}
\end{array}
\right.
\label{eq_stepw}
\end{equation}


\begin{equation}
\stepc(\tage,\taun) = \left\{
\begin{array}{ll}
\frac{(\tage-\taun)^{\betag}}{\taugx^{1+\betag}-\taugi^{1+\betag}}
& \mbox{for $\taun \leq \tage-\taugi$} \\
0 & \mbox{for $\taun \geq \tage-\taugi$}
\end{array}
\right.
\label{eq_stepc}
\end{equation}

\begin{equation}
\tinf= \left\{
\begin{array}{ll}
\tauni & \mbox{for $\tage < \tauni + \taugx(\tauni)$} \\
\taust & \mbox{for $\tage \geq \tauni + \taugx(\tauni)$}
\end{array}
\right.
\label{eq_tinf}
\end{equation}

\acap
and with the following meaning of the symbols:

\begin{itemize}

\item \tauni, \taunx : respectively the nuclear timescales of the most and 
least massive secondary in the SNIa progenitors' binary systems; if \msec\
ranges between 8 and 2 \msun, $\tauni \simeq 0.04$ Gyr and 
$\taunx \simeq 1$ Gyr;
\item \taugi, \taugx : respectively the minimum and maximum gravitational delay
of SNIa DD precursors, originated from systems with given \msec; 
\taugi\ is assumed independent of \msec; \taugx\ is assumed (i) larger 
than the Hubble time for 
all SNIa precursors in the {\em WIDE DD} scheme, (ii) correlated with \taun\
as given by e.g. Eq.(\ref{eq_taugxc}) in the {\em CLOSE DD} scheme;
\item \taust\ is the solution of the equation $\tage-\taun=\taugx(\taun)$, i.e.
the minimum \taun\ which contributes to the explosions at epoch \tage,
which increases with \tage\ (see Fig.~\ref{fig_taunst});
\item \taui\ : the minimum total delay time (\taui=\tauni+\taugi), 
assumed to be independent of \msec;
\item \taux\ : the maximum total delay time, larger than the Hubble 
time for the {\em WIDE DD}; equal to the maximum delay of the least massive
SNIa progenitor for the {\em CLOSE DD}: \taux=\taunx+\taugx(\taunx);
\item \fniii\ : the term describing the dependence on the mass of the DD 
systems, given by Eq.(\ref{eq_fniii});
\item \betaa, \betag\ : the exponents of the power law distributions adopted
respectively for the final separations in the {\em WIDE DD} scheme, and for
the gravitational delays in the {\em CLOSE DD} scheme; a flat distribution of
\aff\ corresponds to $\betaa=0$, and $\betag$ in the vicinity of $-$0.75
(if \mdd\ varies in a small range). 

\end{itemize}


In addition, the ($\alpha,\gamma$) parameters need to be specified, 
in order to compute the distribution of the nuclear delays. 
In the models shown here, I have considered $\alpha=2.35$ and $\gamma=1$:
the dependence of $n(\taun)$ on these parameters is modest for nuclear
delays up to $\sim 1$ Gyr (see Fig. \ref{fig_fiasd}).

In both Eqs (\ref{eq_fiaddw}) and (\ref{eq_fiaddc}) at each total delay \tage,
the \fiadd\ function results from the sum of
the contributions from systems with a range of \msec, and each contribution
is proportional to a power of $(\tage-\taun)$, the latter being just the 
gravitational delay of the progeny of systems born with  secondary mass \msec, 
which end up in a SNIa event at epoch \tage.




The two equations correspond to different characterizations of
SNIa precursors: 
Eq. (\ref{eq_fiaddw}) separately accounts for the sensitivity of
the gravitational delay from the total mass of the DD systems (through the
factor \fniii), and from the distribution function of the separations
(through \betaa). At the same time, it assumes that at any \tage\ up to the 
Hubble time, SNIa explosions come from all systems with  
$\taun \leq \taunx$.
On the contrary, Eq. (\ref{eq_fiaddc}) emphasizes the systematics 
of the range 
of the gravitational delays as \msec\ (and hence \taun) varies: at each \tage, 
only systems with $\taun \geq \taust$ contribute to the explosions, since 
those systems with shorter nuclear timescales have a too short maximum 
gravitational delay.
  
\begin{figure}
\resizebox{\hsize}{!}{\includegraphics{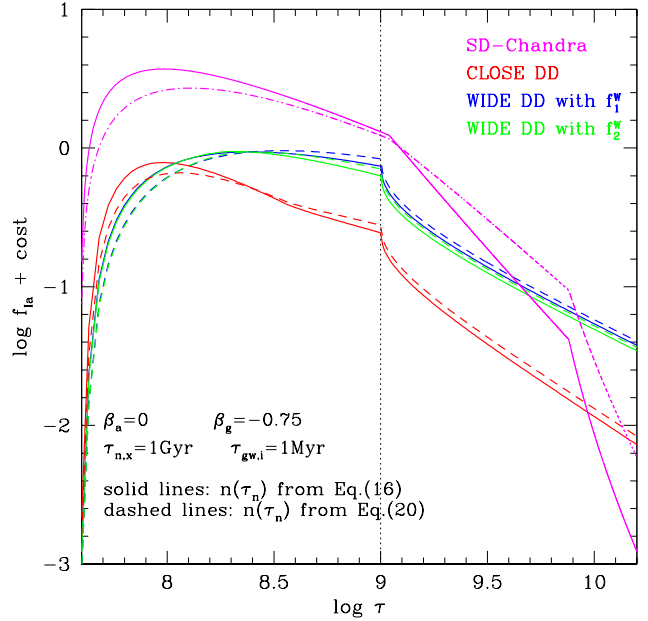}}
\caption{Illustration of the distribution function of the delay times
in the DD model for the labeled values of the parameters. The red lines 
show the result for the {\em CLOSE DD} scheme; the blue and green for the 
{\em WIDE DD} scheme, respectively when \mdd\ is strongly and mildly 
correlated with \msec. Eq.(\ref{eq_taugxc}) has been used for \taugx(\taun).
For comparison,
the magenta curves show the distribution function of the delay times
in the SD model, for Chandrasekhar explosions. In all models the distribution
of the nuclear delays is derived with $\alpha=2.35$ and
$\gamma=1$.} 
\label{fig_fiadd1}
\end{figure}

Neither of the two relations will strictly apply in nature, but they can be 
used to investigate on the general shape of the distribution function of the 
delay times in two extreme situations for what concerns the product of the
Common Envelope evolution. 
As an illustration, Fig. \ref{fig_fiadd1} shows the distribution function
of the delay times for the DD model for one specific choice of the 
parameters, as labeled. 
For comparison, the distribution 
function of the SD systems is
also plotted. I recall here that \fia(\tage) is proportional to the
SNIa rate following an instantaneous burst of SF.

Similar to the case of the SD model, 
the distribution function \fiadd(\tage) appears
characterized  by three regimes: first a rapid increase, followed
by a slow decrease or a wide maximum, and finally a late epoch,
pronounced  decline. This shape can be viewed as a modification of the \fiasd\
function: for the DD model, early explosions are given by systems with
short \taun\ {\em AND} short \taugr; the flat portion corresponds to those
epochs at which the SNIa events come from many combinations
of \taun\ and \taugr; at late epochs we are left with systems with long
\taugr. 
Notice that when \tage\ is large compared to \taunx\, 
Eq.(\ref{eq_fiaddw}) can be approximated 
as $\fiadd(\tage) \propto \tage^{-0.75+0.25\betaa}$ : compared to
the SD model, this decline rate is considerably mild, and basically 
controlled by the dependence of the gravitational
delay on the final separation \aff. A late epoch increase of \fiadd\ can
be realized only if \betaa\ is large and positive,
corresponding to $n(\aff)$ distributions highly skewed toward large
values of the final separations, which is very unlikely.

With respect to the {\em WIDE DD}, 
the {\em CLOSE DD} scheme yields a distribution of the delay times which is
steeper both at the intermediate and at the late epochs. This reflects the
relative paucity of systems with long \taugr\ in this scheme of evolution. 
However, in spite of the very different assumptions, the overall behavior of
the \fia\ functions  in  Fig.~\ref{fig_fiadd1} look similar.
In Sect. 5, the difference between the models will be better
quantified, by considering the distribution function of the
delay times suitably normalized.

For the DD model, at \tage=1 Gyr there is a cusp: mathematically this is 
due to the discontinuity of
the \stepw\ and \stepc\ functions, coupled with the upper limit
of integration for \fiadd. In practice, the cusp occurs at the epoch
at which the systems with smallest \msec\ start contributing to the
SNIa rate, that is \tage=\taunx+\taugi. After this epoch, increasing
\tage\ corresponds to include systems with longer \taugr\, but NOT
longer \taun. The prominence of the cusp is related to the \betaa,(\betag)
exponents which control the $n(\taugr)$ distribution
toward the minimum \taugi.

Finally, the different scheme used to compute the distribution function of
\msec, i.e. whether using Eq.(\ref{eq_phitsc}) or Eq.(\ref{eq_phitgr})
has a very modest impact on the final \fiadd(\tage) function. As for the SD
model, the use of the Greggio \& Renzini (\cite{IO83}) scheme yields a relatively larger rate at late
epochs.

\subsection{Dependence on key parameters.}

I turn now to consider the dependence of the function \fiadd\ on the 
several parameters that need to be specified. 

  
\begin{figure}
\resizebox{\hsize}{!}{\includegraphics{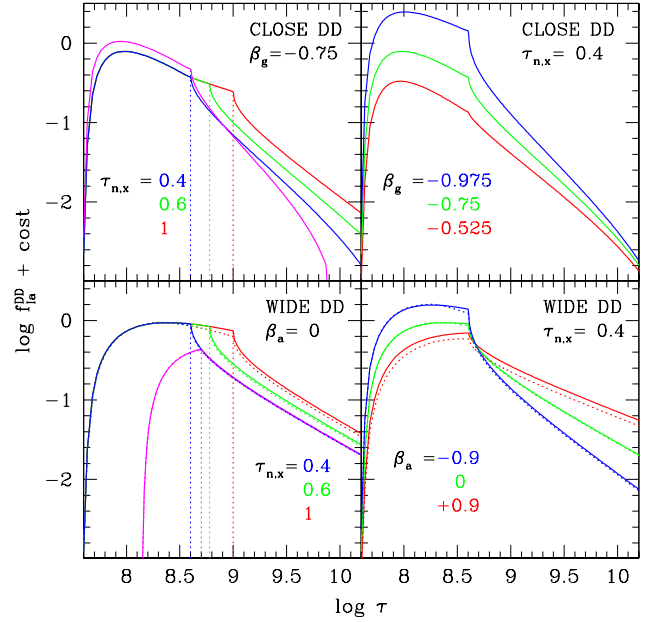}}
\caption{Sensitivity of the distribution function of the delay times
for DD progenitors on various parameters. The top panels refer to the
{\em CLOSE DD} scheme, the bottom panels to the {\em WIDE DD} scheme
of evolution. The left panels show the dependence on the \taunx\ parameter,
with the color encoding as labeled. In addition, the magenta line in the
top panel has been obtained with \taugx(\taun) as given by 
Eq.(\ref{eq_taugxc}) decreased by 0.6 Dex, mimicking the effect of lower \ace.
The magenta line in the bottom panel shows instead the effect of adopting
\taugi=0.1 Gyr in the {\em WIDE DD} scheme. In all other models \taugi=1 Myr
is used. The right panels show the effect of varying \betaa\ and \betag.
The dotted lines in the bottom panels show the results of using \fniii=\fnii
in the computation of \fiaddw.
All curves have been computed with $n(\taun)$ 
derived from Eq.(\ref{eq_phitsc}) with $(\alpha,\gamma)=(2.35,1)$.}
\label{fig_fiadd2}
\end{figure}

Both the upper and the lower limits to the mass of the secondary component
in the SNIa progenitor systems are subject to some uncertainty, which
reflects on the parameters \tauni\ and \taunx. 
The most massive CO DD systems come from progenitors in which both components
are $\sim$ 8 \msun\ stars; if these systems manage to produce a SNIa event, 
\tauni\ is about 0.04 Gyr. As mentioned in the introduction, though, 
the ultimate fate of a double CO WD might be an accretion induced collapse,
rather than a central Carbon deflagration, depending on the modalities
of accretion (e.g. the accretion rate and the angular momentum deposition
on the WD). This question is highly debated in the current literature
(e.g. Piersanti et al. \cite{Amedeo} and Saio and Nomoto \cite{Sano04}),
to the aim of establishing the likelihood of the DD channel as 
SNIa precursors in general. However, notice that 
if the occurrence of the accretion induced collapse depends on the 
mass of the DD components, the DD channel remains a valid SNIa
progenitor, but \tauni\ changes to become the nuclear lifetime of
the most massive secondary in a system which avoids the accretion induced
collapse. Since in the current literature there's no claim of this effect, 
this possibility is neglected here, and all the models adopt 
\tauni = 0.04 Gyr. 
As already mentioned, most double CO WDs come
from systems with \msec\ greater than 2 \msun, so that a reasonable value
for \taunx\ is 1 Gyr. However, the likelihood of a successful explosions
may well be decreasing as \msec\ approaches this limit, due to the
requirement that \mdd\ exceeds the Chandrasekhar mass. Therefore,
\taunx\ is treated as a parameter, and I show here the results obtained
with \taunx=0.4,0.6 and 1 Gyr, corresponding to a lower limit to \msec\
in SNIa progenitors of $\sim$ 3, 2.5 and 2 \msun, respectively.
 
Given the high degree of shrinkage which is obtained when applying the
standard CE recipe, the lower limit to the 
gravitational
delay (\taugi) is likely to be very small. For this reason I adopt here a 
nominal value of \taugi = 0.001 Gyr in most computations. However, the 
minimum gravitational delay could be larger, especially for systems with 
high initial mass ratio, if e.g. Nelemans et al. (\cite{Nele01}) 
scheme of evolution applies. 
Thus, I explore the sensitivity of the results on 
\taugi\ only for the  {\em WIDE DD} scheme, adopting a very large
value of \taugi = 0.1 Gyr. For the {\em CLOSE DD} scheme the critical 
gravitational timescale is instead \taugx: e.g. low values of \ace\ produce
a maximum gravitational delay shorter than the Hubble time even for the lowest
mass systems (see Fig.~\ref{fig_taugx}). Two options for the relation 
\taugx(\taun) are considered.

Without specific population synthesis computations, little can be said 
about the \betaa\ and the \betag\ parameters. Since the distribution of the 
separations in primordial binaries is typically taken to scale as 
$n(\azero) \propto \azero^{-1}$ (e.g. Iben and Tutukov 1984, Han 1998, 
Nelemans et al. 2001), I 
consider the three values $\betaa=-0.9,0,+0.9$ which correspond to assuming 
that, as a result of evolution, the distribution of the separations 
(i) remains basically unchanged;
(ii) flattens off, so that any value of \aff\ is equally probable;
(iii) changes slope, so that more DD systems are found with large \aff.
The values of the \betag\ parameter explored here are related to the
three \betaa\ values via 
$\betag = -0.75+0.25\betaa$ (see Sect. 4.3.2), and are
$\betag = -0.975,0.75,-0.525$. 
Notice that, since most combinations of (\mdd, \aff) lead to short \taugr,
positive values for \betag\ or even
a flat distribution of the gravitational delays are extremely
unlikely.
   

Fig.~\ref{fig_fiadd2} illustrates the dependence of
\fiadd(\tage) on the parameters related to the timescales (left panels),
and on those related to the distribution of the separations (right panels).
The left panels show how the late epoch decline starts at 
\tage=\taunx + \taugi, as argued in the previous section. The more massive
is the lower limit to \msec\ of the SNIa progenitors', the larger is the
fraction of early explosions, i.e. the shorter will be the timescale for 
the release of the bulk of the nucleosynthetic products to the interstellar
medium. If SNIas come mostly from DDs which are born with relatively wide
separations, such that \taugi\ is large, the distribution function
of the delay times behaves like the magenta curve in the lower left panel:
the first explosion occurs at $\tau=\tauni+\taugi$, after which the rate
increases rapidly up to the start of the late epoch decline.
The magenta curve in the top left panel shows instead what happens 
if the close binary evolution produces a large degree of shrinkage:
the \fiadd\ function is more skewed toward short delays, and it 
does not provide systems with delays longer than $\taunx+\taugx(\taunx)$.
This illustrates the potential difficulty of accounting for SNIa in Elliptical 
galaxies, if the close binary evolution produces too close DDs.

The right panels in Fig. \ref{fig_fiadd2} show instead how the distribution
function of the delay times depends on \betaa\ and on \betag\, in the 
special case of \taunx=0.4 Gyr and \taugi=1 Myr. In general, \fiadd\ 
appears to be fairly sensitive to these exponents, with more and more
early explosions as the distribution function of the separations of the DDs
is  more populated at the low \aff\ end.

\section{Discussion}

Very schematically, the distribution function of the delay times of SNIa
progenitors derived in the previous sections for both Single and Double
Degenerate models is characterized by:
\begin{itemize}
\item an early steep rise; 
\item an intermediate phase, hereafter referred to as the wide maximum;
\item a decline phase.
\end{itemize}
The minimum and maximum delay times, the duration of the intermediate
phase, and the slopes of the intermediate and late phases are
different for the Single and the Double Degenerate progenitors, and
are controlled by a few key parameters.

For the SD progenitors, the minimum delay time is equal to the MS
lifetime of the most massive secondary in the primordial binary
producing a SNIa event; the wide maximum phase lasts until a delay
time equal to the MS lifetime of the least massive primary evolving
into a CO WD suitable as SNIa precursor; the decline phase becomes
very steep at late epochs, if the requirement of building up to the
Chandrasekhar mass limits the SNIa progenitors to systems with more
massive primaries in combination with less massive secondaries.  In
the decline phase, the slope of the \fia\ function depends on
the IMF, and on the distribution of the mass ratios.  In particular,
the flatter the distribution of the mass ratios,
the larger the fraction of systems with long delays, at fixed IMF slope. 
For the Chandrasekhar exploders, the
maximum delay is equal to the MS lifetime of the secondary whose
envelope is massive enough to ensure that the most massive CO WD
reaches the Chandrasekhar limit upon accretion. For the Sub--Chandra
exploders, the maximum delay is equal to the MS lifetime of the
secondary with a massive enough envelope to provide the minimum
layer for Helium ignition on top of the companion.  
Both these constraints depend upon the efficiency of the accretion process.


For the DD progenitors, the minimum delay time is equal to the MS
lifetime of the most massive secondary in a SNIa progenitor system,
plus the minimum gravitational delay; the wide maximum phase lasts up
to a delay equal to the MS lifetime of the least massive secondary in
a SNIa progenitor system, again plus the minimum gravitational
delay. The slope of the decline phase is sensitive to the
distribution function of the separation of the DD systems at birth. In
addition, the overall distribution function of the delay times is
steeper if a correlation exists such that the more massive binaries merge 
on a shorter timescale than the less massive ones, due to a more pronounced 
shrinking of the system. This happens in the standard treatment of the 
CE evolution ({\em CLOSE DD} scheme). Finally, the maximum delay time for 
the DDs is basically equal to the
gravitational delay of the least massive and widest DD progenitor: if
the CE stages were to induce a high degree of shrinking, the maximum
delay could well be shorter than the Hubble time.

\subsection {Comparison with the results of populations synthesis codes}

These \fia\ functions have been derived with the aim of providing a general
characterization of the distribution function of the delay times for
the various potential SNIa progenitors, and a number of convenient, though
astrophysically motivated, approximations have been introduced.  It is
thus very important to compare the general shape of these functions
to the results of the population synthesis codes, which follow the
individual evolution of close binaries in detail.  Unfortunately, the
distribution of the delay times of the SNIa events, or equivalently
the SNIa rate following an instantaneous burst of Star Formation (see
Eq. (\ref{eq_rateb})), is not commonly found in the literature. 
Most authors rather quote the current SNIa rate in the Galaxy, which, 
following Eq. (\ref{eq_ratel}),
gives information on the total realization probability of the Ia
channel (i.e. \aia), but not on the shape of the \fia\
function. The most suitable paper to perform a detailed comparison
between the analytic functions presented here and the results of
a population synthesis code is the one by YL.

Fig. 2 in YL shows the SNIa rate following an
instantaneous burst of Star Formation. Four types of precursors
appear in this figure: the DD--Ch, i.e. Chandrasekhar Double Degenerate
exploders;
the SG--Ch, produced by the evolutionary path of the SD--Chandra
considered here; the He--ELD and SG--ELD, which are two flavors of the
Sub--Chandra Single Degenerate channel, the difference being that
the former come from systems with \msec\ greater than
2.5 \msun, which donate Helium to the degenerate companion, while the latter
are systems in which \msec\ is smaller than 2.5 \msun 
\footnote {To be precise, YL quote a secondary mass below (2--3 \msun)},
which donate H, converted to Helium of top of the CO WD.
The He--ELD and SG--ELD can be viewed as two complementary paths building
up into the broad Sub--Chandra category considered here.




The numerical simulations follow the evolution of one single population
of binaries, which evolve through mutually exclusive channels;
the \fia\ functions presented here, instead, are thought of as alternative 
to each other, for the total stellar population. One could consider a scenario
in which SNIa come from different channels,
and construct a composite analytic distribution function of the
delay times by properly assigning the various key parameters, and the
realization probabilities of each channel. I prefer to avoid this approach
and perform the comparison between the results presented here and YL's
by taking into account the different mass ranges which evolve into the
different channels. 

\begin{figure}
\resizebox{\hsize}{!}{\includegraphics{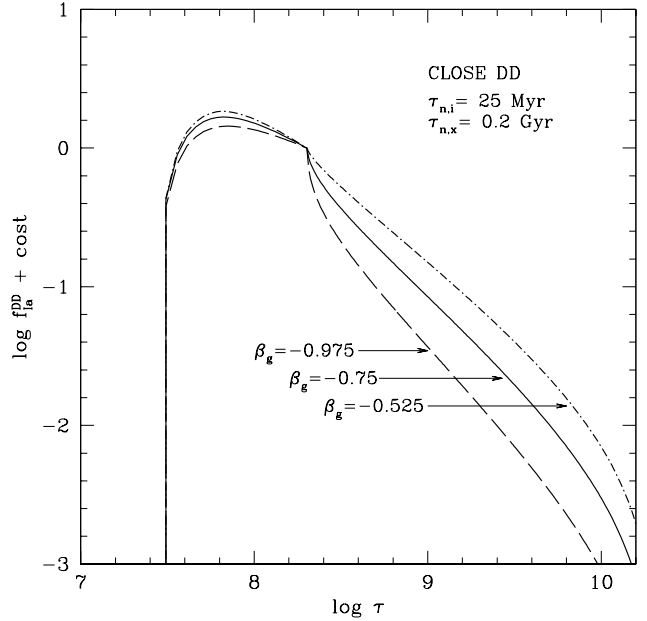}}
\caption{Distribution function of the delay times
for DD progenitors for a choice of parameters which represent
the DD--Ch evolutionary channel in YL. See text for more details.}
\label{fig_fiaddyl}
\end{figure}

Reading off Fig. 2 in YL, the rate for the DD--Ch
exploders starts at a delay time of Log t $\simeq$ 7.4, reaches a maximum
shortly before 0.1 Gyr, and then drops; for delays in excess of about
0.3 Gyr the trend is close to a power law with a slope of $\sim -1.2$.
At 10 Gyr, the rate has dropped of 2.2 Dex with respect to its value
at maximum.
YL state that
in their simulations, the SNIas typically come from binaries with
primary components in the range between 4 and 10 \msun, and that the DD--Ch
channel applies to the systems with secondaries more massive than 4
\msun. The upper limit on \mpri\ is larger than the 8 \msun\ adopted here
because the evolution
in a close binary can prevent C ignition before the loss of the envelope
in stars less massive than $\sim$ 10 \msun\ (Iben \& Tutukov \cite{IT85}). 
The evolutionary lifetime of a 10 \msun\ star (with solar
metallicity) is about 25 Myr, which is also the delay time at which
the first DD--Ch events appear to occur; the lifetime of a 4 \msun\ star 
is about 0.18 Gyr, close to the duration of the peak in the YL
DD--Ch curve.  Since YL adopt a description of the
evolution during both CE phases similar to Eq.~(\ref{eq_ce}), the analogue of
their DD--Ch case would be a {\em CLOSE DD} model with \taui = 0.025 Gyr and
\taunx = 0.18 Gyr, while nothing can be said about the adequate value
of the \betag\ parameter.  Using Eq.~(\ref{eq_taugxc}) with such
short \taunx, the
maximum gravitational delay is much shorter than the Hubble time; on the other
hand, as illustrated in Fig.~\ref{fig_taugx}, \taugx\ is very sensitive
to the various parameters used to describe the CE evolution.
In order to compare the analytic function for the DD model to YL results 
I consider a relation for 
\taugx\ obtained
from  Eq.(\ref{eq_taugxc}) plus a zero point shift of $-$0.6, so as to 
recover maximum delays exceeding the Hubble time for the less massive SNIa
progenitors. This case is shown in 
Fig.~\ref{fig_fiaddyl} for three values of the \betag\ parameter. 
It can be seen that the analytic \fia\ functions are
very similar to the DD--Ch curves in YL; in particular, the case with
\betag=$-$0.75 is very well approximated by a power law with
a slope of $-$1.27 for $\tau \gtrsim 0.5$ Gyr, and 
at 10 Gyr its \fia\ is 2.7 Dex lower than its
maximum value. {\em This is a remarkable similarity, given the
completely different ways in which the two functions have been
obtained}. Notice that in the range $8.7 < \log t < 9.5$ the match 
is better than this, since the analytic curves 
steepen when approaching the maximum delay time.

It is worth to point out that the \fia\ functions for 
the {\em CLOSE DD} scenario are also in broad agreement 
with the SNIa rate following an instantaneous burst of star 
formation for other renditions of the
DD scenario (Tutukov \& Yungelson \cite{Tuyu}; Ruiz-Lapuente
\& Canal \cite{Pilar98}).

The similarity between the analytic and numerical functions for the SD,
Sub--Chandra cases is much less apparent; still the differences can be 
understood in terms of different mass ranges populating this channel.  
After a steep
rise, starting at $\log t = 7.5$ (close to the MS lifetime of an 8 \msun\
star), the rate from the He--ELD channel has a wide maximum, followed
by a rather abrupt drop, which sets in at about 0.5 Gyr. The latter is the
evolutionary lifetime of a $\sim$ 2.7 \msun\ star, not far from 
the least massive secondary populating this
channel, i.e. 2.5 \msun. Systems with lower secondary mass do evolve
on a longer timescale, but do not go through the He--ELD channel.  On
the other hand, the rate from the SG--ELD exploders starts at a delay
time of 0.6 Gyr, in correspondence to the lifetime of the most
massive secondary which evolves through this channel.  The rate rapidly 
reaches a maximum and then
starts declining with a trend close to a power law with a slope of
$-$1.6. At a delay time of $\sim$ 5.6 Gyr the
rate drops dramatically. For comparison, the slope after the wide maximum
of the blue, dot-dashed \fiasd\ curve in Fig.~\ref{fig_fiasd} is 
$\simeq -$1.7. It is tantalizing to conclude that the
analytic formulation presented here for the Sub--Chandra exploders reproduces 
the general trend of the He--ELD + SG--ELD channels in YL, with the late
dramatic drop possibly related to inefficient accretion from low mass
secondaries.

Finally, the rate for the SG--Ch channel is indeed very different from
the analogue analytic \fiasd\ functions in Fig.~\ref{fig_fiasd}. 
However, as stated by YL,
 only systems with \msec\ lower than 2.5 \msun\ evolve along
this path, the more massive secondaries becoming either a DD--Ch, or a
He--ELD. Therefore, the late start of the rate for this channel is
understood in terms of a late \tauni; the rapid drop at delays in
excess of 1 Gyr could instead reflect a low accretion efficiency.
Actually, the comparison of the predictions for the Single Degenerate
Chandrasekhar channel is particularly difficult because of the
different approaches, as anticipated above: the systems evolving through
the SG--Ch channel are clearly a minority in the population synthesis
code.

The general conclusion is that the analytic distribution function of the
delay times presented here can provide an excellent match 
to the results of the population 
synthesis codes, once the appropriate mass ranges and evolutionary
timescales are assumed for the individual evolutionary paths.
Obviously, the numerical simulations also estimate the realization probability
of the various channels, and the total realization probability of the 
Ia event, that is the \aia\ factor. There's no attempt here at evaluating
this factor, which can either be taken from observational estimates,
following Eq. (\ref{eq_ratel}), or from the population synthesis results.

The analytic approach offers several advantages, most notably:
\par\noindent
(i) an easy way to explore the consequences on the evolution
of stellar systems of the various candidates, viewed as alternative SNIa
progenitors; 
\par\noindent  
(ii) it has a built in parametrization of the key properties of
the alternative candidates, i.e. mass ranges, 
IMF and distribution of the mass ratios, distribution of the separations
of the DD systems;
\par\noindent
(iii) a flexible tool to build up an overall distribution
of the delay times by mixing the different 
analytic \fia\ functions, weighted by the relative contribution of the
individual channels to the total realization probability, as astrophysical
considerations may suggest.

\subsection {Comparison between different SNIa candidates}

\begin{figure}
\resizebox{\hsize}{!}{\includegraphics{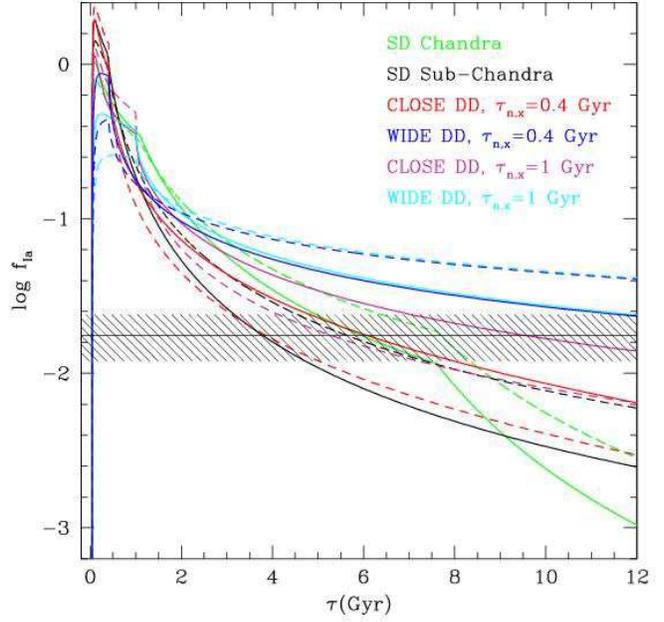}}
\caption{Distribution function of the delay times for SD and 
for DD progenitors as labeled. The plotted functions have been 
normalized to 1 in the range $\tau \leq 12$ Gyr, and the normalization
constant is in units of Gyr$^{-1}$. For the all models the distribution
function of the secondaries has been derived using Eq.(\ref{eq_phitsc}) 
with $\alpha=2.35$. See text for more details.}
\label{fig_fialin}
\end{figure}

For a meaningful comparison of the distribution functions of the delay 
times from the various potential SNIa progenitors it is necessary to
normalize the \fia\ functions. Among the various possibilities, a convenient
normalization is to consider 
$\int_{\taui}^{\taux}\fia(\tage)\cdot \dif \tage = 1 $ (see Sect. 2): 
in this way, 
the specific SNIa rate at the current epoch in a system which formed its
stars in an initial star formation episode of duration \deltat\ is
given by (see Eq. (\ref{eq_ratee})):

\begin{equation}
\frac {\dot{n}_{Ia}^{E}(t)}{\Mstar}= \kalpha \cdot \aia \cdot \avefiab 
\label{eq_rateec}
\end{equation}

\acap
where \avefiab\ is the average of the {\em normalized} distribution 
function of the 
delay times over the age range of the stars in the system, and \aia\ is
the realization probability of the SNIa scenario. I recall here that \kalpha\
depends on the IMF, and is equal to 1.55, 2.83 respectively for Kroupa and
Salpeter IMFs.
Fig.~\ref{fig_fialin} shows some selected cases of the analytic distribution
functions normalized to 1 in the range between 0 and 12 Gyr. Although the
maximum delay for the different candidates plotted here is larger than 
12 Gyr, most distributions steepen enough at late times that
the fraction of SNIa explosions at delay times exceeding 12 Gyr is negligible.
In the attempt of showing a wide range of possible solutions, the
plotted cases adopt the following parameters:
\par\noindent
(i) for the SD models, the solid lines have been computed with $\gamma$=1, 
the dashed lines with $\gamma$=0.005;
the latter choice enhances the relative number of low mass secondaries (see
Fig. \ref{fig_nm2});
\par\noindent
(ii) for the DD models, the solid lines have been computed with \betaa=0,  
\betag=$-0.75$; the dashed lines with \betaa=+0.9, \betag=$-0.975$
for the {\em WIDE DDs} and the {\em CLOSE DDs} respectively. 
The latter cases are meant to
illustrate the wide range of decline rates allowed by these options. 

For a given realization probability of the SNIa event (\aia) all models
plotted in Fig.~\ref{fig_fialin} provide the same total number
of SNIa within 12 Gyr from a burst of SF of given total mass, but the
events are differently distributed in time. Although for all of
the models most of the events occur within the first 1 Gyr,
in the {\em WIDE DD} scheme the explosions are more evenly spread over the
whole 12 Gyr range.
Both the {\em CLOSE DDs} and the SD Sub--Chandra model 
yield an age distribution of the events very skewed at the early
epochs; the SD--Chandra model exhibits a dramatic drop at late epochs, 
related to the requirement of building up to the Chandrasekhar mass by
accreting the envelope of low mass secondaries.

%

The models in Fig.~\ref{fig_fialin} can be compared to the 
observed SNIa rate per unit mass in elliptical galaxies measured by
Mannucci et al. (\cite{Mannu}, hereafter M2005), which is quoted of 
0.044 (+0.016)($-$0.014) SNuM, or 
$0.044 (+0.016)($-$0.014) \times 10^{-3}$ events per \msun\ per Gyr.
If 10\% of the stars in the mass range from 3 to 8 \msun\ end up as SNIa,
the factor $(\kalpha\,\aia)$ is about $(2,3)\times 10^{-3}$ respectively
for Salpeter and Kroupa IMF. Adopting 
$\kalpha\,\aia = 2.5 \times 10^{-3}$, and inserting 
$(\dot{n}_{Ia}^{E}(t)/\Mstar)=0.044 \, 10^{-3}$,
Eq.~(\ref{eq_rateec}) yields 
\avefiab = 0.0176 Gyr$^{-1}$, which is the level indicated by the black line
in Fig.~\ref{fig_fialin}.
%
The dashed region shows the upper and lower limit relative to the range of
the SNIa rate quoted by M2005 for E/S0 galaxies. 
For a larger realization probability of the SNIa event, following either 
from a wider range of progenitor masses, or from a larger probability of the 
SNIa channel within a given mass range, the observational constraint shifts
downward. 

The intercept between the theoretical \fia\ functions and the observational
constraint yields the average age which the stars in early type galaxies 
should have in order to reproduce the data.
So, e.g. the SD--Chandra model is able to fit the data if either
the stellar population in Es is young, or if the realization probability of
the SNIa scenario is larger than what adopted. Both the age of the stars
in Es and the $\kalpha \aia$ factors are uncertain; therefore   
Fig.~\ref{fig_fialin} does not allow us
to draw stringent conclusions about the best model for
SNIa precursors. However, the figure shows the interplay between the various
quantities.

At 12 Gyr the SD models with $\gamma$=1 fall short by about one order of 
magnitude with respect to the level indicated by the observations. 
The mismatch is more severe for the Chandra case. 
Such a big discrepancy is difficult to recover either 
by increasing the realization probability
of the SNIa scenario, or by invoking a younger age for Ellipticals:
on the one hand the black level in Fig.~\ref{fig_fialin} already assumes 
that an important fraction of stars (i.e. $\simeq$ 10 \%) in the suitable mass 
range end up as SNIa. On the other hand, M2005 
data refers to a sample of more than 2000 early type objects, and the
spectrophotometric properties of this class of galaxies strongly
suggest that they are old (see e.g. Renzini \cite{Alvio99}; 
Peebles \cite{Peebles}).
The only possibility to reconcile the SD model with the SNIa rate in
ellipticals seems to be that of assuming a very low $\gamma$ (thin curves),
so that the distribution of the secondaries is maximally populated
at the low mass end. Even so, an accretion efficiency close to 100 \% is 
required for the SD--Chandra models to meet the observations, which seems 
unlikely, as argued in Sect. 3.2.
 
\begin{figure}
\resizebox{\hsize}{!}{\includegraphics{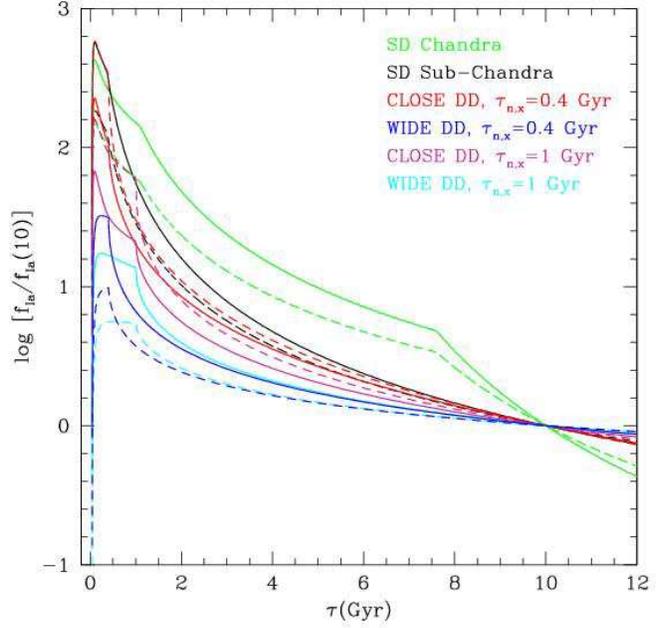}}
\caption{Distribution functions of the delay times for the same models
shown in Fig.~\ref{fig_fialin} with the same color and line coding.
The functions are here normalized to give the same value at
10 Gyr in order to illustrate the different early epoch behavior of models
which reproduce the current SNIa rate in Ellipticals.}
\label{fig_fialin1}
\end{figure}

The DD models more easily account for the observations, provided that  
gravitational delays as long as the Hubble time are realized, i.e. the
common envelope phases do not lead to a too severe shrinking of the DD
systems. The steepest \fia\ function ({\em CLOSE DD} with \taunx=0.4 Gyr and
\betag=$-0.975$) fall short by a factor of $\sim 5$ with respect to the 
observational limit, and are thus unfavored. Notice that the mild slope
of the \fia\ function from intermediate ages onward, implies that assuming 
a younger age for Ellipticals does not efficiently improve the fit for this 
kind of models. The illustration clearly shows that
lower \aia\ and/or older ages for ellipticals are accommodated with  
{\em WIDE DD} models. 

The normalization chosen for Fig.~\ref{fig_fialin} corresponds to assuming that
all models, with the same realization probability, yield the same 
total number of SNIa out of a stellar generation
of unit mass (and therefore the same chemical enrichment). In a 
different approach, Fig.~\ref{fig_fialin1} shows the
models normalized to their value at 10 Gyr: this corresponds to forcing
all of them to fit the current SNIa rate in Es with an average 
stellar age of 10 Gyr, albeit with different values for the factor 
$\kalpha \aia$. The realization probabilities required by this normalization
are of the order of $10^{-2}$ for the SD models, of $10^{-3}$ for the DD 
models, but there is a noticeable dependence of the \aia\ factors on the
various parameters defining the models, including the formalism
to describe the binary population (i.e. whether Eq. (\ref{eq_phitsc}) or 
Eq. (\ref{eq_phitgr}) are used).
Only the SD-Chandra models with low $\epsilon$ < do require a totally
unrealistic realization probability, corresponding to $\ga 100 \%$
of the stars with mass between 2 and 8 \msun, and will not be 
considered further.
Once normalized in this way, the various models correspond to dramatically 
different evolution over cosmic time of the SNIa rate from a burst 
of SF (notice that the rate is plotted on a logarithmic scale). 
This property offers an important tool to 
discriminate among the models by looking at the impact on the large
scales, like the Iron Mass--to--Light ratio in Clusters
of galaxies, the evolution with redshift of the SNIa rate in
Ellipticals, and the systematic trend of the SNIa rate with galaxy 
type (Greggio \cite{IO05}). 
The latter point is addressed in the following section.

\begin{figure}
\resizebox{\hsize}{!}{\includegraphics{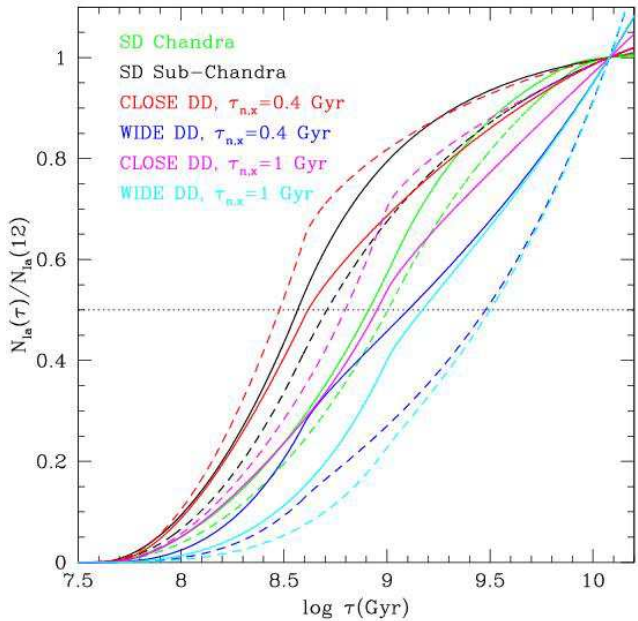}}
\caption{Cumulative number of SNIa explosions following an instantaneous
burst of SF for the same models shown in Fig.~\ref{fig_fialin},
with the same color and line coding. In particular, the thin blue and cyan
lines are especially flat {\em WIDE DD} cases; the thin red and magenta lines 
are especially steep {\em CLOSE DD} models.}
\label{fig_nia}
\end{figure}

To conclude this section, Fig.~\ref{fig_nia} shows the cumulative 
number of SNIa explosions
as a function of age for an instantaneous burst of SF, normalized to
the total number of events within 12 Gyr.
This figure illustrates another distinctive 
characteristic of the various SNIa models: the time scale over
which 50 \% of the total SNIa explosions from an instantaneous burst of SF 
have occurred.
This timescale can be taken as indicative of the typical delay
with which the Fe from SNIa is released to the Interstellar Medium, thereby
decreasing its $\alpha$/Fe ratio. Within the explored range of models
such timescales varies between 0.3 and 3 Gyr going from the steepest 
{\em CLOSE DD} case, to the flattest 
{\em WIDE DD} case. The SD--Chandra model with $\gamma$=0.005 
(green dashed line)
has a typical timescale of $\simeq$ 1 Gyr, which is often taken as a
reference value to infer the formation timescale of systems  
exhibiting an $\alpha/{\rm Fe}$ overabundance. Fig.~\ref{fig_nia} 
emphasizes that such timescale {\em does}  
depends on the SNIa model, namely it is longer the flatter the
distribution of the delay times of the SNIa progenitors is. 
The estimated formation timescales, then, remain uncertain by a factor of 
a few, modulo the actual SNIa channel that dominates in nature.

\subsection {SNIa rate in different galaxy types}

As anticipated in the previous section, the large differences of
the SNIa rate temporal behavior typical of the various  
models translate into a different trend of the SNIa rate as a function of 
the galaxy type, thereby offering a tool to discriminate among the potential
progenitors. This has already been outlined in Sect. 2, where the
Cappellaro et al. (\cite{Capp99}) data have been shown to indicate that 
the ratio between the \fia\ value at late delay times and its average
value over the whole range of delay times (up to the Hubble time) 
should be $\simeq 0.15$.
For the analytic functions, the quantity 
$\frac{\langle f_{\rm Ia}\rangle_{10,12}}{\langle f_{\rm Ia}\rangle_{0,12}}$
%
is equal to 0.02 (0.03), 0.05 (0.08)  respectively for the 
SD Chandra (Sub--Chandra) models with $\gamma=1$ and 0.005 ; 
equal to 0.07 and 0.15 for the {\em CLOSE DDs} with \taunx=0.4,1 
(and \betag=$-$ 0.75); 
while it is $\simeq$ 0.23 for the {\em WIDE DDs} with a flat 
distribution of the separations (\betaa=0). Therefore, the ratio between 
the SNIa rate
in Ellipticals and Spirals indicates that the Single Degenerate model
underestimates the current rate in early type galaxies, with respect
to the rate in late types, a result of its fast decline at late times. 

This constraint has been derived considering a schematic description of the
star formation history in early and late type galaxies, and by using 
a theoretical value for the $\Mstar/L_{\rm B}$ ratio in the two galaxy types.
A much better constraint on the SNIa model progenitors can be built upon
the recent results by M2005, by considering the 
trend of 
the SNIa rate {\em per unit galaxy mass} with the parent galaxy type. 
In fact, Eq.~(\ref{eq_rate}) can be written as:
  
\begin{equation}
\frac{\dot{n}_{Ia}(t)}{\Mstar} = \kalpha \cdot \aia \cdot \avefiapsi 
\label{eq_ratepsi}
\end{equation}

\par\noindent
where \avefiapsi\ is the average of the distribution function of
the delay times weighted by the star formation rate over the galaxy
lifetime, and where  the integral of the SFR has been approximated with the
galaxy (stellar mass) \Mstar\,\footnote {As pointed out in Sect. 2, 
the actual stellar mass in a galaxy can be $\approx$ 30 (45) \% lower than 
the integrated SFR for Salpeter (Kroupa) IMF, due to the mass return 
form dying stars.}. Eq.~(\ref{eq_ratepsi}) clearly shows that
the trend of the SNIa rate with galaxy type reflects the systematics
of the their star formation histories. Given that \fia\ is a decreasing
function of the delay time, the younger systems will have a higher
SNIa rate per unit mass, by an amount which depends on  
the shape of the \fia\ function. 
The correlation in M2005 
can be interpreted
as the result of Eq. (\ref{eq_ratepsi}), where the different SF histories
imply both a different SNIa rate per unit mass, and a different $B-K$
color.

\begin{figure*}
\centering
\includegraphics[width=17cm]{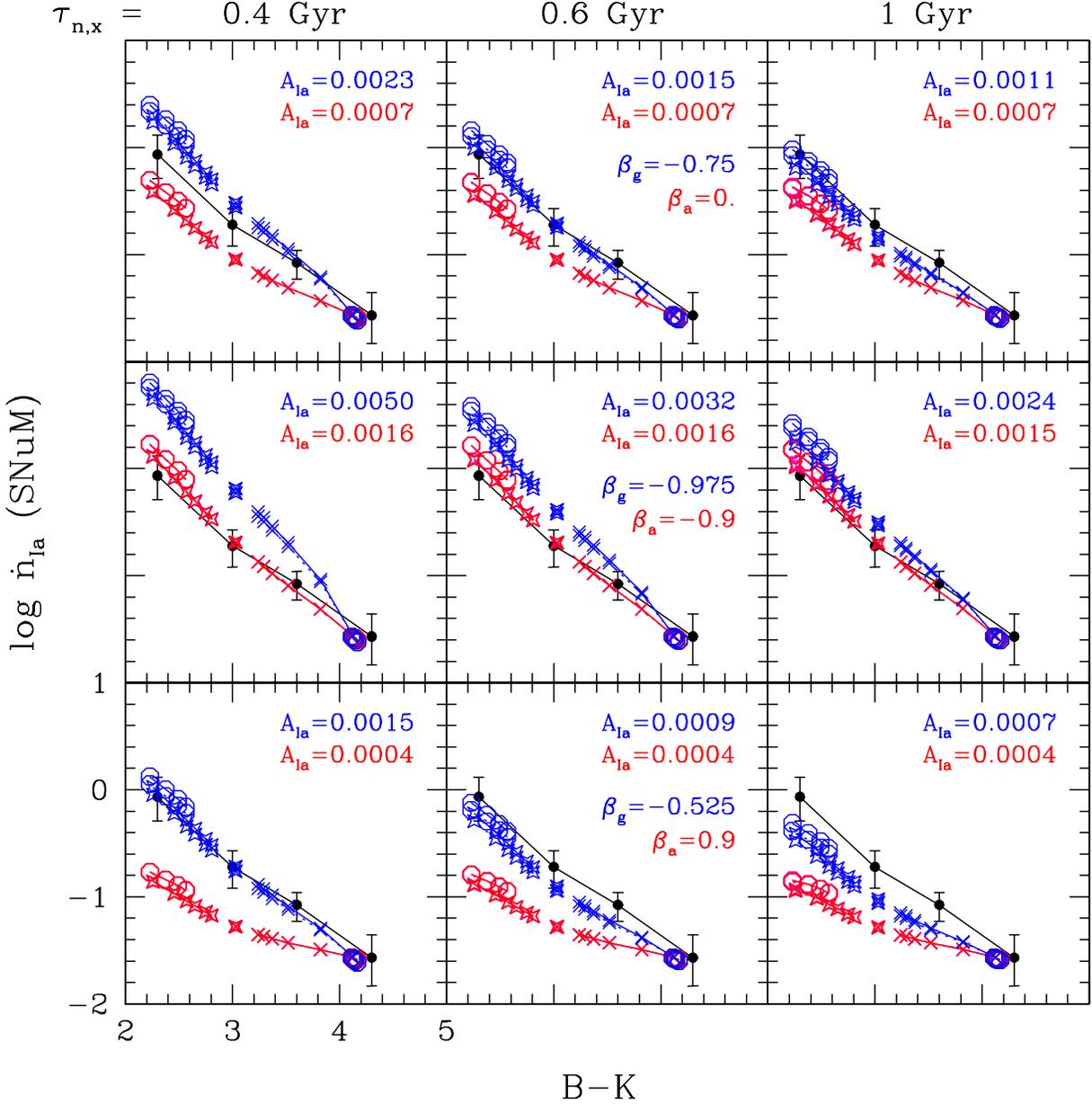}
\caption{Comparison between Double Degenerate model predictions and 
observations of the SNIa rate per unit mass as a function of the color 
of the parent galaxy. The data from M2005 paper are plotted
as black dots, with their quoted error bars, and connected with a solid
line. The 9 panels show the trend with $B-K$ as a tracer of SF history 
(see text) of {\em CLOSE DD} (blue) and {\em WIDE DD} (red) models: 
from left to right the minimum secondary
mass in SNIa progenitor systems decreases, and its MS lifetime \taunx\ 
increases; from top to bottom the slope in the decline phase varies as 
indicated in the central panels, reflecting different distribution 
functions of the DD separations. 
Also labeled are the realization probabilities
required by the different models in order to reproduce the current rate
in the reddest galaxies.
All models adopt $\alpha=2.35$ and $\gamma=1$, and both options 
for the distribution function of the secondaries, i.e. Eq.(\ref{eq_phitsc}) and
Eq.(\ref{eq_phitgr}) are displayed. The minimum gravitational delay is 
assumed to be negligible; the maximum gravitational delay for the 
CLOSE DDs follows Eq. (\ref{eq_taugxc}).
The two options for the WIDE DD case, i.e. \fniii=\fni and \fnii in 
Eq. (\ref{eq_stepw}) give the same results, and indeed are both plotted in
of the various panels.}
\label{fig_mandd}
\end{figure*}
\begin{figure}
\resizebox{\hsize}{!}{\includegraphics{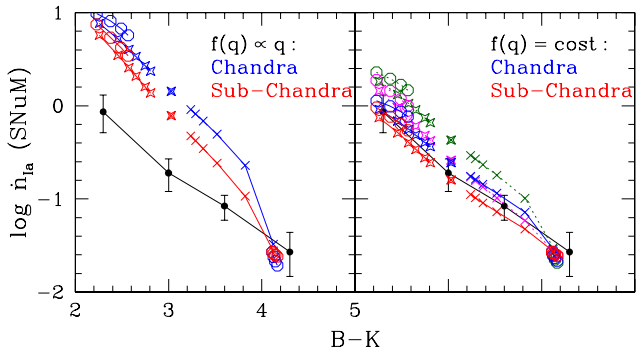}}
\caption{Comparison between Single Degenerate model predictions and 
M2005. The point type encodes the SFH, as in Fig.~\ref{fig_mandd}. 
All models adopt $\alpha=2.35$; other parameters are labeled.
The blue and red curves refer to models computed with $n(\msec)$ given by
Eq.(\ref{eq_phitsc}); the green and magenta loci show the results 
obtained with Eq.(\ref{eq_phitgr}) and $\gamma=1$.}
\label{fig_mansd}
\end{figure}

Figures \ref{fig_mandd} and \ref{fig_mansd} show the M2005 
observed correlation, and
the theoretical predictions for the various SNIa models, computed as
follows.
To describe the SF history in the various galaxy types I have considered
for four families of models:

\par\noindent
(1) old burst models, characterized by a constant SFR within a starting
epoch $t=0$ and ending epoch $t_{\rm B} = 0.5, 1, 1.5$ and 2 Gyr;
\par\noindent
(2) exponentially decreasing models, all starting at epoch $t=0$ and 
with different e--folding times $t_{\rm SF}$:
\acap
$\psi(t) = {\rm e}^{-\frac{t}{t_{\rm SF}}}$
with $t_{\rm SF} = 1,3,5,7,9,11$ Gyr;
\par\noindent
(3) exponentially increasing models: all starting at epoch $t=0$ and 
with a different characteristic time $\tau_{\star}$:
\acap
$\psi(t) = {\rm e}^{-\frac{12-t}{\tau_{\star}}}$
with $\tau_{\star} = 8,6,4,3,2,1$ Gyr. 
\acap
These models correspond to a sequence of age
distributions all peaked at 12 Gyr, and decreasing e-folding age, so that
the lower $\tau_\star$ the younger the average age of the stellar 
population;
\par\noindent
(4) young burst models, characterized by a constant SFR within a starting epoch
$t_\star = 9.5, 10, 10.5, 11$ Gyr and all ending at $t=12$ Gyr.

\par\noindent
In addition, a model with a constant SFR over the whole range of 12 Gyr 
has been computed.
The considered range of SF histories 
appears to encompass the range of $B-K$ colors of the galaxies in
M2005 sample. The composite colors have been computed using 
Girardi et al. (2000) simple stellar population models with solar metallicity. 
The bluest and reddest bin in M2005 are 
defined as lower and upper limits only (i.e. $B-K<2.6$ and $B-K>4.1$); here 
they have been specified so that they are close to the bluest and 
the reddest model stellar population. 
The point type in Figs~\ref{fig_mandd} and \ref{fig_mansd} encodes the 
SF history: burst models
are plotted as circles; exponential models as crosses, each family (2 and 3 
described above) connected by a line. The solitary dot shows the
model with a constant SF rate over 12 Gyr.

The theoretical rates have been normalized so as to reproduce
the observed values in the reddest galaxies with a SFH given by the 
old burst model with a duration of 2 Gyr;
consequently, the normalization implies
different realization probabilities of the SNIa event for the different 
models.
The values for \aia\ required by this normalization, having
adopted a Kroupa IMF 
and Eq. (\ref{eq_phitsc}) for
the distribution function of the secondaries, are labeled in the
figure. \aia\ is the number fraction of SNIa events out of a stellar
generation, and should be compared to the number fraction of stars in
a mass range which can lead to the event, e.g. $3 \leq m/\msun \leq 8$.
For Kroupa IMF, 20 \% of the stars fall in this mass range, and therefore
the normalization requires that in the {\em WIDE DD} scenario, approximately
3 \% of the stars in the suitable mass range
should end up as SNIa.
For the {\em CLOSE DD} scenario the analogous fraction is around $\sim$ 10 \% .
The dependence of these figures on the parameters for the DD model can be
appreciated from the values labeled in Fig.~\ref{fig_mandd};
for Salpeter IMF, the required fraction of SNIa events from the same mass
range is smaller by a factor of $\sim$ 0.6.
It can be noticed that choosing any other
old burst model, or the oldest exponentially decreasing model, 
{\em would not change the normalization}.
 
It appears that the both families of DD models do fit well the observational 
data when the choice of the parameters is such to provide an intermediate
shape of the distribution function of the delay times. 
The {\em CLOSE DD}s yield a too steep evolution of the SNIa rate per unit mass
with galaxy color, for a low \betag\ and a short \taunx; similarly, the
{\em WIDE DD}s give a too flat relation if \betaa\ is large, as well as \taunx.
This comparison does not necessarily favor the {\em WIDE} or the 
{\em CLOSE DD}s; rather it points to a $moderate$ solution: either relatively
flat {\em CLOSE DD}s or relatively steep {\em WIDE DD}s. 

Fig.~\ref{fig_mansd} shows the analogous results for the SD
models. The left panel clearly shows that, with the standard choice of
the parameters (i.e. $n(\msec)$ as in the population synthesis computations,
$\alpha=2.35$ and $\gamma=1$) both Chandra and Sub--Chandra
exploders imply a too large increase of the SNIa rate per unit mass going from
early to late type galaxies. The right panel shows the results for a choice
of the parameters aimed at improving the match between the SD model and 
the observations. 
When using Eq.~(\ref{eq_phitgr}) to describe $n(\msec)$ the agreement is
better (green and magenta points), but is seems that  
only with a very 
low value of $\gamma$ (i.e. if all the mass ratios $q=\msec/\mpri$ are equally
probable) can the SD model be reconciled with the observations. 
In this case, the normalization to the rate in the reddest 
galaxies implies 
that $\sim$ 10\% of all stars born with $2\leq m/\msun \leq 8$ should end up
as SNIa of the SD Chandra variety; alternatively $\sim$ 12\% 
of all stars born with $3\leq m/\msun \leq 8$ should end up as SD Sub--Chandra
SNIa. 

The comparison between the M2005 data and the theoretical models 
does not definitely rule out any of the alternative progenitors,
but puts constraints on the key parameters within the various families.
The precise values of these parameters are subject to some uncertainty.
M2005 convert their observed SNIa rate per unit luminosity into
a rate per unit mass by using the results from the galaxy models by Bell \&
de Jong (\cite{Bell01}), which are much more complicated than the SF histories 
considered here. This introduces a (small) inconsistency between the models
and the data. Another caveat concerns the approximation of the galaxy mass
with the integrated SF rate, which affects the zero point (by not more
than a factor of 2), and the slope of the theoretical trend. 
Both these approximations, however, hardly affect the major conclusion 
that indicates that
\par\noindent 
(i) the distribution function of the delay times should drop by a factor of
$\approx$ 100 from its early peak to its value at 10 Gyr;
\par\noindent
(ii) in principle all candidates can reproduce the observed trend of the
SNIa rate per unit mass with galaxy color, with an adequate combination
of the key parameters. However, the SD channel seem
to require quite some fine tuning; in particular the Chandra model
is viable only if the accretion efficiency is very large, which 
appears unlikely.

\section{Summary and concluding remarks}

In this paper I have presented a straightforward formalism to relate the
rate of SNIa events in stellar systems to their star formation history 
through two fundamental characteristic of the SNIa progenitor model: the 
realization probability of the SNIa scenario from a single age stellar
population \aia,
and the distribution function
of the delay times \fia(\tage), which is
proportional to the SNIa rate past an instantaneous burst of star 
formation.

The various models for the SNIa progenitors correspond to different values for
\aia, and to different shapes of the \fia\ function. No attempt is
made here to give a theoretical value for \aia, which can be derived either 
from the numerical
realizations in population synthesis models, or directly from the observations.
The latter suggest $\aia$ on the order of $10^{-3}$ (see Sect. 2.1). A more
precise estimate could be derived with a detailed modeling of the
star formation in late type galaxies, and with a robust assessment of the
IMF.
For the \fia\ function instead, the paper presents
analytical formulations to describe the most popular SNIa progenitor models. 
It is shown that some parameters play a key role
in shaping the distribution function of the delay times, most
notably: the mass range of the secondaries in systems which provide SNIa
events; the minimum mass of the primary which yields a massive enough
CO WD to ensure the explosion, and the efficiency of accretion on top of it,
for the SD model; the distribution of the separations of the 
DD systems at their birth. In addition, it is found that the algorithm 
adopted to describe the
distribution of the stellar masses in binaries has an important impact on
the slope of the \fia\ function for the SD model.

In all of the three scenarios (SD Chandra and Sub--Chandra, and DD Chandra)
the distribution function of the delay times is characterized by a steep,
early rise, so that the maximum rate is reached soon after the first
SNIa explodes. 
For the SD model, this reflects the behavior
of the {\it clock} ($\dot{\msec}$) which describes how the rate of
change of the evolutionary mass decreases as the delay time
increases. The quantity $|\dot{\msec}|$ drops so fast to
prevail over the increase of the number of SNIa precursors 
as the delay time increases. For the DD model, the {\it clock} has an
additional contribution from the gravitational delay; still the early
steep rise is present because most DD systems have a short \taugr,
given that a wide range of masses and separations correspond to 
a short gravitational delay. 
Therefore, the property of the distribution function of the delay times 
to reach its maximum shortly after the first event occurs has a very
robust justification. As repeatedly said, following an instantaneous burst of
SF, the epoch of the first event is equal to 
the MS lifetime of the most massive secondary (\tauni), for the SD model, 
plus the minimum gravitational delay (\taugi), for the DD model.
There's no apparent reason to consider an upper limit for \msec\ in SNIa
progenitor systems smaller than the most massive primary which provides a
CO WD (i.e. an $\sim$ 8 \msun\ star). This implies $\tauni \simeq 0.04$ Gyr.
As shown in Fig.~\ref{fig_shrink}, even in the {\em WIDE DD} scheme
the binary WDs can emerge from the CE with a very small separation, so that
also \taugi\ is likely to be very short. Therefore, in the context of
the binary evolution, it seems very hard to accommodate a shape for
the distribution of the delay times similar to the one that according to
Strolger et al. (\cite{Strol04}) best explain the redshift dependence of the 
SNIa rates measured in GOODs, i.e. a Gaussian distribution centered on
$\tage \sim 3-4$ Gyr. Such a distribution could be obtained if 
SNIa events were produced only by systems with \msec\ in
the vicinity of 1.3 \msun,  for the SD model, or by DD systems born with
a separation of about 2.5 \rsun\ (but notice that a little spread of the
binary mass implies anyway a large spread of delays). Both 
hypothesis are very contrived. Furthermore,
a Gaussian distribution would not meet the constraint on the \fia\ function
derived from the trend of the SNIa rate with parent galaxy type, like in
M2005. 

The analytical functions derived in this paper compare very well
to the results of the population synthesis codes in the 
literature, and especially to those in Yungelson and Livio (2000), 
when taking into account the appropriate mass ranges and kind of progenitor. 
The Monte Carlo simulations follow the evolution of a population
of binaries, which evolve through mutually exclusive channels, according to 
the response of the system to the mass exchange phases.
The remarkable similarity between the analytic functions derived here and
the numerical results suggests that the shape of \fia\ is mainly
determined by the mass ranges and general characteristics of the clock
of the explosion, while the 
details of the response of the individual systems to the RLO events are of
lesser importance. 

Once put on the same formalism, the various models for the SNIa progenitors
can be compared to the relevant observations in order to judge which one 
best accounts for the data. The most direct observational counterpart of the
\fia\ functions is the redshift dependence of the SNIa rate in elliptical
galaxies (per unit galaxy mass), due to the fact that the bulk of 
their stars formed at very early epoch, and the single burst is a 
reasonable approximation. Along this line, 
Maoz \& Gal-Yam (\cite{Maoz04}) have compared the SNIa
rate in galaxy clusters at redshift between 0 and 1 to the family of \fia\
functions proposed by Madau et al. (\cite{Piero98}), reaching the 
conclusion that the average
delay time of SNIa precursors ought to be $\le 2$ Gyr.
The notion of a typical delay time for the SNIa precursors has very little 
justification in the context of stellar evolution in binaries, since 
a wide range of delay times is produced by any kind of progenitor.
Rather, the slow increase of the SNIa rate with redshift reported in
Maoz \& Gal-Yam (\cite{Maoz04}) would favor 
the DD model, perhaps better of the {\em WIDE DD} variety. With the progress
of the Cluster SN surveys (see e.g. Maoz (\cite{Maoz05})) we will be able 
to further investigate the redshift dependence of the rate of SNIa in
Elliptical galaxies, and come to stronger conclusions. 
A less direct, but definitely complementary and effective approach to 
constrain to SNIa progenitor model is attempted here, by considering the 
predicted trend of the SNIa rate (per unit mass) with the parent galaxy type 
(see also Della Valle \& Livio \cite{Dvalle94}; Ruiz-Lapuente, 
Burkert \& Canal \cite{Pilar95}). 
The M2005 data suggest that the DD channel is
favored with respect to the SD channel, and that the distribution of the
separations of the DD systems should be such to produce a moderate decline
of the \fia\ function at late times. The SD model is not completely ruled out
by this comparison, but it requires 
(i) a flat distribution of the mass ratios and
(ii) accretion efficiencies 
%
close to 100 \%. This can be accomplished only if the matter 
is accreted and burned on top of the WD at the same pace, so as to avoid 
either expansion beyond the Roche Lobe 
(and the formation of a Common Envelope), or the accumulation of a 
Hydrogen layer which is eventually ignited under degenerate conditions (so that
a Nova explosion occurs). Even in the Hachisu et al. (\cite{HKN96}) scenario
part of the accreted matter is lost by the system in a stellar wind.

The various models for the SNIa progenitors have different impact on the
large scales; 
some preliminary considerations
are in Greggio (\cite{IO05}), while more detailed investigations
will be presented elsewhere. Here I just remark a few points. 

Once normalized to reproduce the 
current SNIa rate in Ellipticals, the SD model corresponds to a large Fe mass
to light ratio in Cluster of Galaxies (see Fig. 13, and Greggio \cite{IO05}). 
A detailed study of the expected Fe mass to light ratio in galaxy clusters
as a function of the various possible progenitors, and its evolution 
with redshift  will allow us to better constrain the SNIa model and the 
contribution of SNIa to the Fe in the intracluster medium. 

The timescale over which, following an instantaneous burst of SF, 
half of the Fe is released to the interstellar medium varies between 
$\sim$ 0.3 to 3 Gyr for the wide variety of SNIa progenitor models considered
here. Accordingly, the formation timescale of systems which exhibit 
an enhancement of $\alpha$ elements with respect to Fe is rather uncertain,
and depends on the SNIa model.
Quantitatively, the actual constraint on the formation timescales also
varies with the duration of the star formation episode in the system
(Matteucci \& Recchi \cite{MR01}).
Preliminary computations show that, in a star forming system, such timescale
may range between 1 and several Gyr. 
This problem will be discussed in a forthcoming paper.

The evolution  of the gas flows in 
Ellipticals depends on the balance between the rate of mass return and 
the SNIa rate, past a burst of SF. 
The former scales with time as $\propto t^{-1.3}$, while, at delay times
greater than $\sim 1$ Gyr, the \fia\ functions scale as $\propto t^{-s}$ with
$s$ 
$\sim -1$ for the {\em DD WIDE}, $\sim -1.2$ 
for the {\em DD CLOSE},
$\sim -1.6$ for the SD Sub--Chandra. Therefore, it appears that the  
secular evolution of the SNIa rate past a burst of SF is critically
close to the evolution of the rate of mass return, and that
the fate of the gas in Ellipticals is very sensitive to the SNIa 
progenitor's model. It is also worth noticing that the shape of the 
\fia\ function is different from
a simple power law, as is adopted in Ciotti et al. (\cite{Ciotti}) to 
model the gas flows in ellipticals. In particular, the presence of the wide 
maximum phase at intermediate epochs will impact on the dynamical evolution
of the gas. 

In this paper, the emphasis has been put on the intercomparison of 
the various models for the SNIa progenitor. Actually, all different 
channels could 
contribute to the SNIa events, each with its own probability, as in the
realizations of the population synthesis models. Some diversities of the
observational properties of SNIa have been found in the literature, which
support this notion  (e.g. Branch \cite{Branch}, 
Benetti et al. \cite{Benetti}). 
In particular the different luminosity at maximum, 
and the different decline rate of the light curve,  
as measured by the the $\Delta m_{15}$ parameter of Phillips (\cite{Phillips}),
of the events in early and late type galaxies (Della Valle \& 
Panagia \cite{Dvalle92}, van den Bergh \& Pazder \cite{vdB}, 
Hamuy et al. \cite{Hamuy}, Garnavich \& Gallagher \cite{Garnavich}) 
could be related to different typical progenitors. 
If both the Single and Double Degenerate channels are at work with similar 
total realization
probabilities, in early type galaxies the DD explosions should
prevail over SD events, since the distribution function of the delay times
of the latter declines fast. In late type galaxies, instead, all channels
should contribute to the current rate, with a larger proportion of SNIa from
the SD channel, due to their high rate at early epochs.
The formalism presented in this paper allows a straightforward
exploration of the effect of a mixture of progenitors, e.g. 
by varying the relative \aia\ realization probabilities.
Eventually, it will be possible
to constrain the mixture of progenitors by modeling the evolution of
the SNIa rate in galaxies of different types, and considering as well
all the other consequences on the large scales. 

\begin{acknowledgements}
I am indebted to Alvio Renzini for a critical reading of the manuscript and
many useful suggestions. I also thank Luca Ciotti for discussions on the
mathematical formalism, and Francesca Matteucci and Simone Recchi for
having renewed my interest in the problem of SNIa theoretical rates.
This work was partly supported by the Italian Ministery of University and
Research (MURST) under the grant COFIN 2003. 
\end{acknowledgements}

\appendix

\section{Equations used to derive the \fiadd(\tage) function}

\subsection{An approximate relation for the gravitational delay}
\label{appe_fmr}

The gravitational delay time is given by:

\begin{equation}
\taugr = \frac {0.15 \aff^4}{(\mpwd+\mswd)\mpwd\mswd} {\rm Gyr}.
\label{eqa_taugr}
\end{equation}

\acap
The mass dependent term can be written as

\[\fmr = \mdd^2 \cdot \mswd - \mswd^2 \cdot \mdd \] 

\begin{figure}
\resizebox{\hsize}{!}{\includegraphics{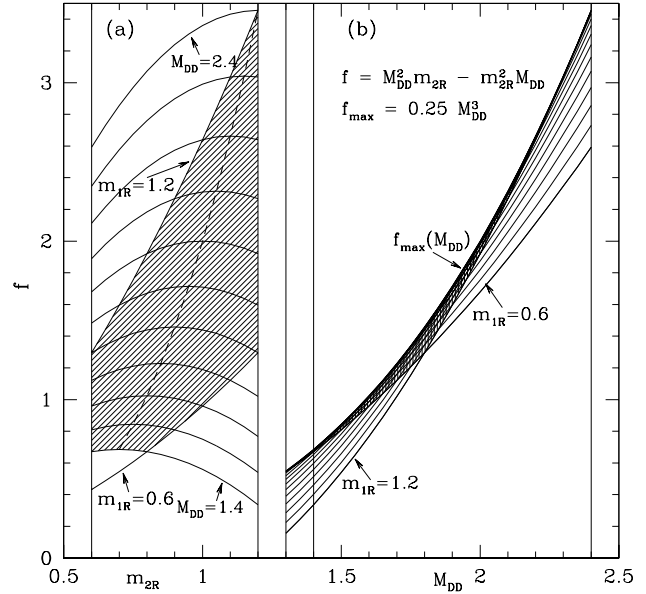}}
\caption{Mass dependent term of \taugr\ plotted versus \mswd\ for
selected values of \mdd\ (panel a), and plotted versus \mdd\ for selected 
values of \mswd\ (panel b). The dashed curve in panel (a) connects the 
maxima of the parabolas. Along this line \mswd=\mpwd. The shaded area 
delimits the parameter space of SNIa producers.}
\label{fig_fmr}
\end{figure}

\acap
with \mdd=\mpwd+\mswd , i.e. 
the total mass of the DD system. Given that WD masses range between 0.6 and
1.2 \msun, \mdd\ goes from 1.2 to 2.4 \msun.
\fmr\ is a family of parabolas, both viewed as a function of \mswd, and of
\mdd. These families are shown in Fig. (\ref{fig_fmr}).
Each parabola in panel (a) has an absolute maximum in $\mswd=0.5\mdd$, of
$\fmrx=0.25\cdot\mdd^3$. This locus is drawn in panel (a) as the
dashed line, and in panel (b) as the uppermost thick curve.
Not all combinations of (\mdd,\mswd) 
are acceptable, and the shaded area in the two panels show the allowed
parameter space. In addition to the vertical limits 
$1.4 \leq \mdd/\msun \leq 2.4$, and $0.6 \leq \mswd/\msun \leq 1.2$,
it is further required that $\mpwd(=\mdd-\mswd)$ ranges between 0.6 \msun\ and 
1.2 \msun. The more restrictive criterion $\mswd \leq \mpwd \leq 1.2 \msun$
is unnecessary. Notice that, occasionally, the evolution in Close Binaries
produces $\mpwd<\mswd$ (even though, by definition, $\mpri \geq \msec$),
e.g. in some cases of conservative RLO when the secondary may become more
massive than the original primary, thereafter leaving a more massive WD 
remnant. 
 
The loci \mpwd=0.6\msun\ and \mpwd=1.2\msun\ 
are shown in Fig.~\ref{fig_fmr}:
along each parabola, only the portion included between
the two loci is acceptable. Basically, at given \mdd\ we exclude those
values of \mswd\ which imply \mpwd $<$ 0.6 \msun, which is a He WD; similarly, 
we exclude those values of \mswd\ which imply \mpwd $>$ 1.2 \msun, taken here
as the upper limit to the mass of a WD.

As clearly visible in panel b) of Fig.~\ref{fig_fmr}, 
the term \fmr\ is much more sensitive to \mdd\ rather than to \mswd, and
there is an almost one to one correspondence between \mdd\ and \fmr\ over
the whole parameter space of double CO WDs.
In the computations presented here I approximate the function \fmr\ 
with its maximum value of  $0.25 \cdot \mdd^3$, which is the upper envelope
of the family of parabolas in panel (b). This approximation leads
to Eq. (\ref{eq_taugrap}). 

\subsection{Shrinkage following the RLO}
\label{appe_shrink}

Given the uncertainty of the results of the Common Envelope
evolution I have considered two alternatives. In one case, both mass transfer
phases are regulated through the standard CE Eq. (\ref{eq_ce}), which results
into:

\begin{equation}
\frac{\af}{\ai} = 0.5\ace\, \frac{\mdf\,m}{\mdi}\, \left( \frac{\mdi-\mdf}
{r_{\rm L}} + 0.5\,\ace\,m \right)^{-1} 
\label{eqa_sce}
\end{equation}

\acap
where \md\ indicates the mass of the donor, $A$ indicates the separation
of the system, and the subscripts $i$ and $f$ refer to before and after
the RLO, respectively. Further, $m$ is the mass of the 
companion, and $r_{\rm L}$ is the Roche Lobe radius in units of the initial 
separation, which is adopted from Eggleton (1998), and is generally 
$\sim 0.4-0.5$. 
The blue and cyan dots in Fig. \ref{fig_shrink} result from the twofold 
application of this equation, first with

\mdi=\fen $\cdot$ \mpri, \, $m$=\msec, \, and \, \mdf=\mpwd\ as given by 
Eq.~(\ref{eq_mwdc});

\acap
second with

\mdi=\fen $\cdot$ \msec, \, $m$=\mpwd, and \, \mdf=\mswd\ as given by 
Eq.~(\ref{eq_mwdc}).

In the alternative evolutionary scheme, the first mass transfer is considered 
parametrized by the {\em Envelope Ejection} relation in Nelemans et al. (\cite{Nele01}), reported here 
as Eq.~({\ref{eq_ne}). Since, in the same notation adopted above:

\begin{displaymath}
 J_{\rm i}=\mdi m \sqrt{\frac{G\ai}{\mdi + m}}\,\,\, , \,\,\, 
   J_{\rm f}=\mdf m \sqrt{\frac{G\af}{\mdf + m}} 
\end{displaymath}

\begin{displaymath}
  \Delta M = \mdi - \mdf \,\,\, , \,\,\, M_{\rm B} = \mdi + m. 
\end{displaymath}
 
\acap
it follows

\begin{equation}
\frac{\af}{\ai} = \left( \frac{\mdi}{\mdf} \right)^2 \, \frac{\mdf+m}{\mdi+m}
\, \left( 1-\gamma\frac{\mdi-\mdf}{\mdi+m} \right)^2.
\label{eqa_ne}
\end{equation}

Eq. (\ref{eqa_ne}) is applicable if

$$\gamma \frac{\mdi-\mdf}{\mdi+m} \leq 1$$  

\acap
otherwise the system would loose more than 100\% of its original angular 
momentum. With a little algebra, the last relation becomes:

$$\frac{m}{\mdi} \geq \tilde{q} = \gamma \left( 1-\frac{\mdf}{\mdi} \right)-1$$

\acap
that is a lower limit to the mass ratio of the system at RLO. Typically,
the quantity (\mdf/\mdi) (which is the core mass fraction of the donor 
at RLO) ranges
between 0.15 to 0.25 for \mdi\ between 8 and 2 \msun. Then,
$\tilde{q} \simeq 0.2,0.4,0.6$ for $\gamma=1.5,1.75,2$. 
Close to this lower limit, the \afai\ ratio from Eq. (\ref{eqa_ne}) becomes 
very small, and indeed smaller than that given by Eq. (\ref{eqa_sce}). 

The red and magenta points in Fig. \ref{fig_shrink} are generated with the 
following prescriptions: similar to Nelemans et al. (\cite{Nele01}), 
the first mass transfer results into a shrinkage given by  
Eq. (\ref{eqa_sce}) if the mass ratio
is smaller than $\tilde{q}$, or the maximum \afai\ from  Eq. (\ref{eqa_sce}) 
and Eq. (\ref{eqa_ne}) if the mass ratio is larger than $\tilde{q}$.
At the second mass transfer, Eq. (\ref{eqa_sce}) is applied. 

The product of the two \afai\ ratios resulting from the first and the second
mass transfer phases naturally equals the ratio between the final and the
original separation of the close binary (\affai) plotted in 
Fig. \ref{fig_shrink}. 

\subsection{Function $g(\tage,\taun)$ for WIDE DDs}
\label{appe_gnele}

In this section I derive an expression for the fraction of systems which,
having a nuclear delay \taun, have a total delay smaller than \tage\ for
the {\em WIDE DD} evolutionary scenario.
Under the assumptions that the total binary mass (\mdd) and the  
separation
of the DD system (\aff) are independent variables, the contribution to 
gravitational delay \taugr\ from systems with separations in the range
(\aff,\,\aff+d\aff) is:

\begin{equation}
\dif n(\taugr,\aff) = n(\aff)\, n(\mdd)\, \dif \aff\, \dif \mdd
\end{equation}

\acap
where $n(\aff)$ and $n(\mdd)$ are the distribution functions of the separations
and total binary mass of the progeny of systems born with a secondary mass 
whose nuclear timescale is equal to \taun, and the variables combine so that
\taugr=0.6$\frac{\aff^4}{\mdd^3}$. It follows:

\begin{equation}
n(\taugr)\dif \taugr = \dif \taugr \int_{\affn}^{\affx}{n(\aff) \, n(\mdd) \,
\left| \frac{\partial \mdd}{\partial \taugr} \right| \, \dif \aff}
\label{eqa_ntgrne0}
\end{equation}

\acap
where \affn\ and \affx\ define\ the range of separations which 
lead to the same gravitational delay 
\taugr. Given the relation between gravitational delay, DD mass and
separation, \affn\ and \affx\ are directly related to the  
minimum and maximum \mdd:

\begin{displaymath}
  \affn=\frac{\mddn^{0.75}\taugr^{0.25}}{0.6^{0.25}}\,\,\, , \,\,\,
   \affx=\frac{\mddx^{0.75}\taugr^{0.25}}{0.6^{0.25}}.
\end{displaymath}

\begin{figure}
\resizebox{\hsize}{!}{\includegraphics{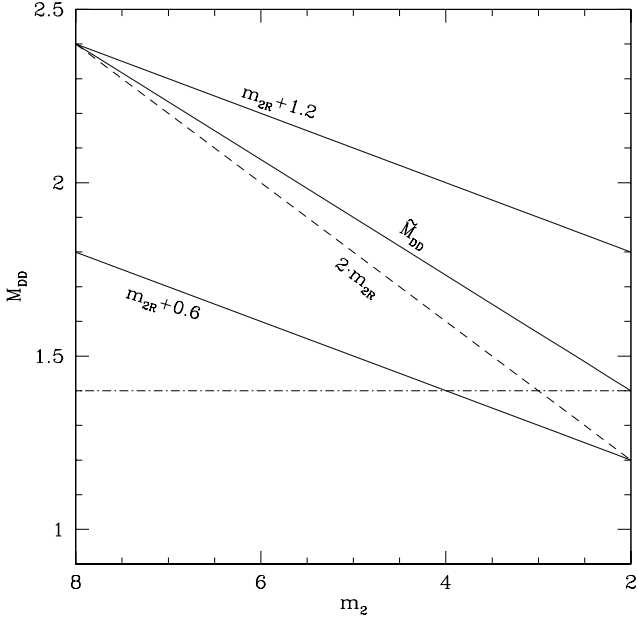}}
\caption{Limits for the total mass of the DD systems produced by
primordial binaries with secondary mass \msec, leaving a remnant \mswd.
The line labeled \mddt\ is used in the computations to explore the effect
of a tight correlation between \msec\ and its progeny \mdd.}
\label{fig_mddlim}
\end{figure}

Fig.~\ref{fig_mddlim} shows plausible limits for \mdd, illustrating that 
the heavier \msec\ is, the heavier its \mdd\ progeny. 
To proceed, formulations for $n(\aff)$ and $n(\mdd)$ need to be specified:
since the {\em WIDE DD} case is meant to describe a situation in which 
the evolution produces DDs in a wide range of separations, and
since the relevant range of final separations is rather narrow (from
0.5 to 4.5 \rsun), as a convenient parametrization I adopt:

\begin{equation}
n(\aff) \propto \aff^{\betaa}.
\end{equation}

\acap
As for the distribution of the DD masses, I explore the consequences of
two extreme assumptions:

\acap
(1) all values are equally probable: \, \, \,$n(\mdd) =$ constant within

$$\mddn=\max (1.4,\mswd+0.6);\, \, \, \mddx=\mswd+1.2$$

\acap 
with \mswd\ given by Eq. (\ref{eq_mwdc});   

\acap
(2) the distribution is peaked around the value\footnote{This arbitrary 
relation is used just to explore the
effect of a systematic decrease of \mdd\ as \msec\ decreases.}
(see Fig.~\ref{fig_mddlim}):

$$\mddt=1.4+(\msec-2)/6.$$

\acap
With respect to case (1), notice that, although one could 
enforce $\mddn=2\mswd$, there are evolutionary paths
which could lead to $\mswd>\mpwd$, e.g. when after the Ist RLO the 
separation is larger than the primordial one. 
\par\noindent
Case (2) corresponds to assuming
that there is a tight correspondence between \msec\ and the mass of the DD
system, so that, at any \msec\ (i.e. \taun), \mdd\ is specified, and
the distribution of the gravitational delays of systems
with given \taun\ is: 

\begin{displaymath}
n(\taugr) \dif \taugr \propto n(\aff) \dif \aff \,\,\,\,\, 
\mbox{with $\aff= (\mddt)^{0.75}\taugr^{0.25}/0.6^{0.25}$} .
\end{displaymath}

With some algebra, Eq (\ref{eqa_ntgrne0}) yields (for $\betaa \neq -1$):

\begin{equation}
n(\taugr) \propto \fniii \cdot \taugr^{0.25\betaa-0.75}  
\label{eqa_ntgrne}
\end{equation}

\acap
where

\begin{equation}
\fni \propto 
\mddt^{0.75+0.75\betaa} 
\label{eqa_fni}
\end{equation}

\acap
for a narrow distribution of \mdd\ around \mddt, while

\begin{equation}
\fnii \propto \mddx^{1.75+0.75\betaa} - \mddn^{1.75+0.75\betaa} 
\label{eqa_fnii}
\end{equation}

\acap
for a wide distribution of \mdd.

\begin{figure}
\resizebox{\hsize}{!}{\includegraphics{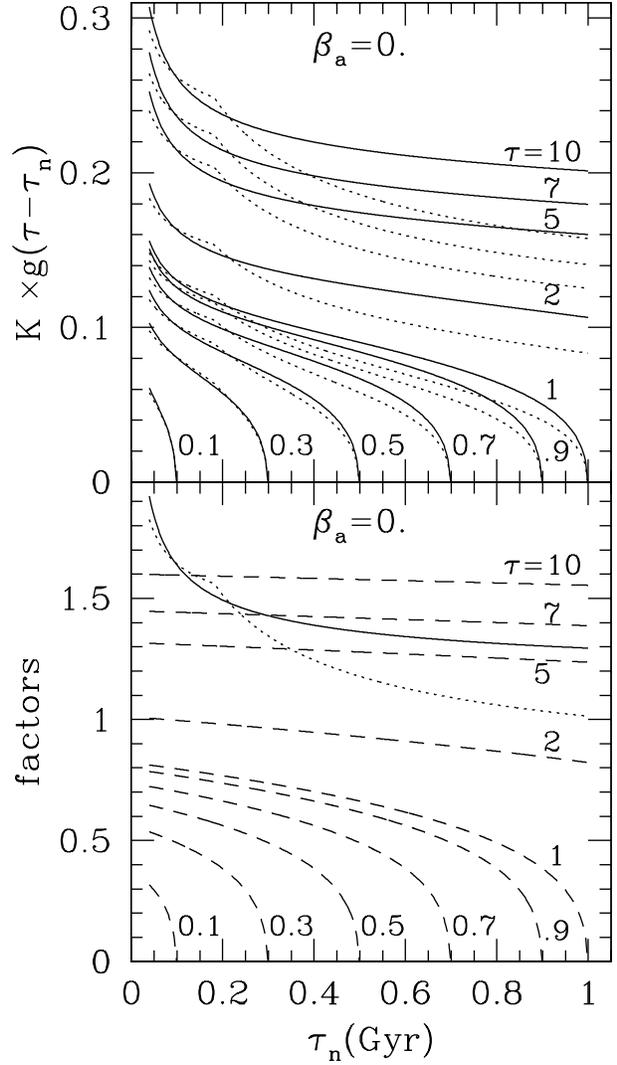}}
\caption{Lower panel: factors in the RHS of Eq.(\ref{eqa_gttn}) for a flat 
distribution of \aff. The dashed lines show the \tage\ dependent term and
are labeled with the \tage\ value in Gyr. Notice how, as \tage\ increases
toward \taunx=1 Gyr the baseline on \taun\ increases, while for $\tage > 1$
the curves just shift upward.
Upper panel: resulting $g$ function in arbitrary units. Solid (dotted) lines 
result from using Eq.(\ref{eqa_gttn}) with \fni (\fnii).}
\label{fig_gfun}
\end{figure}


Inserting Eq. (\ref{eqa_ntgrne}) in Eq. (\ref{eq_gttn0}),
the $g(\tage,\taun)$ function is derived as:

\begin{equation}
g(\tage,\taun) \propto \left\{
\begin{array}{ll}
0 & \mbox{for \tage $\leq \tage_1$} \\
\fniii \,\left[ (\tage-\taun)^{\betaat}-\taugi^{\betaat})
\right]
& \mbox{for $\tage_1 \leq \tage \leq \tage_2$}
\end{array}
\label{eqa_gttn}
\right.
\end{equation}

\acap
with $\betaat = 0.25(1+\betaa)$, $\tage_1 = \taun+\taugi$ , 
$\tage_2 = \taun+\taugx$, and 
where the third branch has been neglected because it is assumed
assume that the maximum gravitational delay
is larger than the Hubble time for every \taun.


The number of systems which have a nuclear delay \taun\ and a total delay 
up to \tage\ is proportional to two factors: one (\fniii) describing the
systematics with \mdd, the other scaling with the width of
the parameter space in \taugr. The dependence on the distribution of the
separations appears in the exponents of both factors.
The lower panel of Fig. ~\ref{fig_gfun} illustrates these two factors 
for \betaa=0, which corresponds to a flat distribution of 
final separations. The \tage\ dependent quantity  
\footnote{A small value of \taugi=0.001 Gyr
has been used in this figure}
(dashed lines) is larger when the 
total delay is larger: more systems merge {\em within} a longer total delay.
At fixed \tage, this factor decreases for increasing \taun, as the
available range in \taugr\ decreases. At long 
total delay times this effect becomes less important, since the upper limit to
\taun\ is of 1 Gyr only.
The solid and the dotted lines in the lower panel of Fig.~\ref{fig_gfun}
show respectively the \fni\ and \fnii\ factors, which account for
the systematics of the gravitational delay with the nuclear delay: 
longer \taun\ correspond to lower \msec, and then to less massive \mdd.
At fixed \aff, systems with lower \mdd\ are diluted over a larger 
\taugr\ range: this effect is reflected on the decreasing \fniii\ factors
with increasing \taun, and is more pronounced for \fnii\ because of the 
additional
systematics with the range $(\mddx-\mddn)$, which gets smaller as \taun\ 
increases (see Fig.~\ref{fig_mddlim}).
 
The upper panel in Fig. ~\ref{fig_gfun} shows the resulting (non-normalized
\footnote{A proper definition requires $g(\taugx+\taun,\taun)=1$ at 
every \taun, which is a relation between \taugx\ and \taun. Since for the
{\em WIDE DD} scheme we consider $\tage < \taugx$ 
for every \taun\ the normalization factor is not important.}) 
function $g(\tage,\taun)$ for
the two considered distributions of \mdd. By construction, the $g$ functions
are zero for $\taun \geq \tage-\taugi$ (see Eq. (\ref{eqa_gttn})). The fraction
of systems within a given total delay time \tage\ decreases as 
\taun\ increases, a dependence which is more pronounced when the variation 
in the range of \mdd\ with \msec\ is accounted for (i.e. when using \fnii).
At any \taun\, the fraction of systems with a delay time smaller than \tage\ 
increases with \tage. The variation of $g$ with \taun\ and \tage\ 
depends on the
distribution function of the final separations \aff, as discussed
in Sect.4.3 and illustrated in Fig.~\ref{fig_gfun1}.

\subsection{The differential distribution function of the delay times}
\label{appe_leib}

It its generic form, the Leibniz integral rule is: 

\begin{displaymath}
\frac{\dif}{\dif z}\int_{a(z)}^{b(z)}f(z,x) \dif x=
\int_{a(z)}^{b(z)}\frac{\partial f}{\partial z} \dif x \, 
+f(z,b)\frac{\dif b}
{\dif z} \, - f(z,a)\frac{\dif a}{\dif z}. 
\end{displaymath}

\acap
This can be applied to Eq. (\ref{eq_fiadd0}), with 

\begin{displaymath}
z = \tage \hspace{0.5cm} x = \taun \hspace{0.5cm} f(z,x) = n(\taun)\cdot g(\tage,\taun)  
\end{displaymath}
\begin{displaymath}
a = \tauni \hspace{1cm} b = \min(\taux,\tage).
\end{displaymath}

\acap
Let's consider in turn the three additive terms at the right hand side 
of the Leibniz rule. The last term is equal to:

\[-[n(\taun)g(\tage,\taun)]_{\taun=\tauni}\,\,\frac{\dif \tauni}
{\dif \tage}=0\]

\acap
because \tauni\ is constant.

\acap
The second term is equal to:

\[[n(\taun)g(\tage,\taun)]_{\taun=\min(\taunx,\tage)}\,\,
\frac{\dif \min(\taunx,\tage)}{\dif \tage}=0\]

\acap 
because 

\acap
(i) in $\tage \ge \taunx$:\,\,\, $\frac{\dif \taunx}{\dif \tage}=0$ 
\,\,\, since \taunx\ is constant;

\acap
(ii) in $\tage \leq \taunx$:\,\,\, $g(\tage,\taun=\tage)=0$
\,\,\, by construction, since the $g$ function is null in 
$\taun \geq \tage-\taugi$ (see Eq. (\ref{eq_gttn0})).

\acap
Therefore, only the first term is left:

\begin{equation}
\fiadd(\tage) = \int_{\tauni}^{\min(\taunx,\tage)} n(\taun) \, 
\frac{\partial g}{\partial \tage} \, \dif \taun .
\label{eqa_fiadd}
\end{equation}

\acap
The $g(\tage,\taun)$ function is continuous (see Eq.~\ref{eq_gttn0}) and 
thus its derivative can be computed in all its points. However, the function 
presents cusps in \tage=\taun+\taugi\ and in \tage=\taun+\taugx, where the
derivative will not be continuous.

\acap
Eq. (\ref{eqa_gttn}) ({\em WIDE DDs}) can be written as:

\begin{displaymath}
g(\tage,\taun) \propto \left\{
\begin{array}{ll}
0 & \mbox{for $\taun\geq\tage-\taugi$} \\
\fniii [(\tage-\taun)^{0.25(1+\betaa)}-\taugi^{0.25(1+\betaa)}]
& \mbox{for $\taun\leq\tage-\taugi$}
\end{array}
\right.
\end{displaymath}

\acap
for delay times up to the Hubble time.
Upon derivation this equation yields:

\begin{equation}
\frac{\partial g}{\partial \tage} \propto \left\{
\begin{array}{ll}
\fniii \,(\tage-\taun)^{-0.75+0.25\betaa}
& \mbox{for $\taun \leq \tage-\taugi$} \\
0 & \mbox{for $\taun \geq \tage-\taugi$} .
\end{array}
\right.
\label{eqa_dgn}
\end{equation}

Inserting Eq.(\ref{eqa_dgn}) into Eq.(\ref{eqa_fiadd}) the
distribution function of the delay times for the double degenerate
systems in the {\em WIDE DD} evolutionary scheme is obtained as:

\begin{equation}
\fiadd(\tage) \propto \int_{\tauni}^{\min(\taunx,\tage)}n(\taun) \,
\stepw(\tage,\taun) \, \dif \taun 
\label{eqa_fiaddw}
\end{equation} 

\acap
with

\begin{displaymath}
\stepw(\tage,\taun) = \left\{
\begin{array}{ll}
\fniii \,(\tage-\taun)^{-0.75+0.25\betaa}
& \mbox{for $\taun \leq \tage-\taugi$} \\
0 & \mbox{for $\taun \geq \tage-\taugi$}
\end{array}
\right.
\end{displaymath}

\acap
For the {\em CLOSE DD} formulation, it's convenient to notice that the third
branch in Eq. (\ref{eq_gttc}) requires $\tau \ge \taun + \taugx(\taun)$,
and therefore exists only if the total delay time considered is sufficiently
long: $\tage \geq \tauni+\taugx(\tauni)$. Therefore, I re-write 
Eq. (\ref{eq_gttc}) as:

\acap
$-$ if $\tage \leq \tauni+\taugx(\tauni)$:

\begin{equation}
g(\tage,\taun) = \left\{
\begin{array}{ll}
\frac{(\tage-\taun)^{1+\betag}-\taugi^{1+\betag}}{\taugx^{1+\betag}-\taugi^{1+\betag}} & \mbox{for $\taun \leq \tage-\taugi$}\\ 
0 & \mbox{for $\taun \geq \tage-\taugi$}
\end{array}
\right.
\label{eqa_gttc1a}
\end{equation}

\acap
$-$ if $\tage \geq \tauni+\taunx(\tauni)$:

\begin{equation}
g(\tage,\taun) = \left\{
\begin{array}{lll}
1 & \mbox{for $\taun \leq \taust$} \\
\frac{(\tage-\taun)^{1+\betag}-\taugi^{1+\betag}}{\taugx^{1+\betag}-\taugi^{1+\betag}} & \mbox{for $\taust \leq \taun \leq \tage-\taugi$}\\ 
0 & \mbox{for $\taun \geq \tage-\taugi$}
\end{array}
\right.
\label{eqa_gttc1b}
\end{equation}

\acap
where \taust\ is the solution of the equation $\tage=\taun+\taugx(\taun)$.
The derivative is then:

\acap
$-$ if $\tage \leq \tauni+\taugx(\tauni)$:

\begin{displaymath}
\frac{\partial g}{\partial \tage} \propto \left\{
\begin{array}{ll}
\frac{(\tage-\taun)^{\betag}}{\taugx^{1+\betag}-\taugi^{1+\betag}} 
& \mbox{for $\taun \leq \tage-\taugi$} \\
0 & \mbox{for $\taun \geq \tage-\taugi$} .
\end{array}
\right.
\end{displaymath}

\acap
$-$ if $\tage \geq \tauni+\taunx(\tauni)$:

\begin{displaymath}
\frac{\partial g}{\partial \tage} \propto \left\{
\begin{array}{lll}
0 & \mbox{for $\taun \leq \taust$} \\
\frac{(\tage-\taun)^{\betag}}{\taugx^{1+\betag}-\taugi^{1+\betag}} & \mbox{for $\taust \leq \taun \leq \tage-\taugi$}\\ 
0 & \mbox{for $\taun \geq \tage-\taugi$}
\end{array}
\right.
\end{displaymath}

\acap
The last two equations can be written in a compact form as:

\begin{displaymath}
\frac{\partial g}{\partial \tage} \propto \left\{
\begin{array}{ll}
\fddc(\tage,\taun) & \mbox{for $\taun \leq \tage-\taugi$} \\
0 & \mbox{for $\taun \geq \tage-\taugi$} .
\end{array}
\right.
\end{displaymath}

\acap
with

\begin{displaymath}
\fddc(\tage,\taun) = \left\{
\begin{array}{ll}
\mbox{for $\tage < \tauni+\taugx(\tauni)$:} & \hspace{-3cm} 
\frac{(\tage-\taun)^{\betag}}{\taugx^{1+\betag}-\taugi^{1+\betag}}\\
\mbox{for $\tage \geq \tauni+\taugx(\tauni)$:} \left\{
\begin{array}{ll}
0 & \mbox{for $\taun \leq \taust$}\\
\frac{(\tage-\taun)^{\betag}}{\taugx^{1+\betag}-\taugi^{1+\betag}} & 
\mbox{for $\taun \geq \taust$}
\end{array}
\right.
\end{array}
\right.
\end{displaymath}

Inserting the last two equations 
into Eq. (\ref{eqa_fiadd}), and splitting the
integration over \taun\ in two parts, one from \tauni\ to \taust\ and
the other from \taust\ to the upper limit ($\min(\taunx,\tage)$), the 
distribution of the delay times is obtained as:

\begin{equation}
\fiadd(\tage) = \int_{\tinf}^{\min(\taunx,\tage)}n(\taun) \,
\stepc(\tage,\taun) \, \dif \taun
\label{eqa_fiaddc}
\end{equation} 

\acap
with

\begin{equation}
\stepc(\tage,\taun) = \left\{
\begin{array}{ll}
\frac{(\tage-\taun)^{\betag}}{\taugx^{1+\betag}-\taugi^{1+\betag}}
& \mbox{for $\taun \leq \tage-\taugi$} \\
0 & \mbox{for $\taun \geq \tage-\taugi$}
\end{array}
\right.
\label{eqa_stepc}
\end{equation}

\acap
and

\begin{equation}
\tinf= \left\{
\begin{array}{ll}
\tauni & \mbox{for $\tage < \tauni + \taugx(\tauni)$} \\
\taust & \mbox{for $\tage \geq \tauni + \taugx(\tauni)$}
\end{array}
\right.
\label{eqa_tinf}
\end{equation}

\end{document}